\def\PsfigVersion{1.10}
\def\setDriver{\DvipsDriver} 
\ifx\undefined\psfig\else \fi
\let\LaTeXAtSign=\@
\let\@=\relax
\edef\psfigRestoreAt{\catcode`\@=\number\catcode`@\relax}
\catcode`\@=11\relax
\newwrite\@unused
\def\ps@typeout#1{{\let\protect\string\immediate\write\@unused{#1}}}

\def\DvipsDriver{
	\ps@typeout{psfig/tex \PsfigVersion -dvips}
\def\PsfigSpecials{\DvipsSpecials} 	\def\ps@dir{/}
\def\ps@predir{} }
\def\OzTeXDriver{
	\ps@typeout{psfig/tex \PsfigVersion -oztex}
	\def\PsfigSpecials{\OzTeXSpecials}
	\def\ps@dir{:}
	\def\ps@predir{:}
	\catcode`\^^J=5
}

\def\figurepath{./:}

\def\DoPaths#1{\expandafter\EachPath#1\stoplist}
\def\leer{}
\def\EachPath#1:#2\stoplist{
  \ExistsFile{#1}{\SearchedFile}
  \ifx#2\leer
  \else
    \expandafter\EachPath#2\stoplist
  \fi}
\def\ps@dir{/}
\def\ExistsFile#1#2{%
   \openin1=\ps@predir#1\ps@dir#2
   \ifeof1
       \closein1
   \else
       \closein1
        \ifx\ps@founddir\leer
           \edef\ps@founddir{#1}
        \fi
   \fi}
\def\get@dir#1{%
  \def\ps@founddir{}
  \def\SearchedFile{#1}
  \DoPaths\figurepath
}

\def\@nnil{\@nil}
\def\@empty{}
\def\@psdonoop#1\@@#2#3{}
\def\@psdo#1:=#2\do#3{\edef\@psdotmp{#2}\ifx\@psdotmp\@empty \else
    \expandafter\@psdoloop#2,\@nil,\@nil\@@#1{#3}\fi}
\def\@psdoloop#1,#2,#3\@@#4#5{\def#4{#1}\ifx #4\@nnil \else
       #5\def#4{#2}\ifx #4\@nnil \else#5\@ipsdoloop #3\@@#4{#5}\fi\fi}
\def\@ipsdoloop#1,#2\@@#3#4{\def#3{#1}\ifx #3\@nnil 
       \let\@nextwhile=\@psdonoop \else
      #4\relax\let\@nextwhile=\@ipsdoloop\fi\@nextwhile#2\@@#3{#4}}
\def\@tpsdo#1:=#2\do#3{\xdef\@psdotmp{#2}\ifx\@psdotmp\@empty \else
    \@tpsdoloop#2\@nil\@nil\@@#1{#3}\fi}
\def\@tpsdoloop#1#2\@@#3#4{\def#3{#1}\ifx #3\@nnil 
       \let\@nextwhile=\@psdonoop \else
      #4\relax\let\@nextwhile=\@tpsdoloop\fi\@nextwhile#2\@@#3{#4}}
\ifx\undefined\fbox
\newdimen\fboxrule
\newdimen\fboxsep
\newdimen\ps@tempdima
\newbox\ps@tempboxa
\long\def\fbox#1{\leavevmode\setbox\ps@tempboxa\hbox{#1}\ps@tempdima\fboxrule
    \advance\ps@tempdima \fboxsep \advance\ps@tempdima \dp\ps@tempboxa
   \hbox{\lower \ps@tempdima\hbox
  {\vbox{\hrule height \fboxrule
          \hbox{\vrule width \fboxrule \hskip\fboxsep
          \vbox{\vskip\fboxsep \box\ps@tempboxa\vskip\fboxsep}\hskip 
                 \fboxsep\vrule width \fboxrule}
                 \hrule height \fboxrule}}}}
\fi
\newread\ps@stream
\newif\ifnot@eof       
\newif\if@noisy        
\newif\if@atend        
\newif\if@psfile       
{\catcode`\%=12\global\gdef\epsf@start{
\def\epsf@PS{PS}
\def\epsf@getbb#1{%
\openin\ps@stream=\ps@predir#1
\ifeof\ps@stream\ps@typeout{Error, File #1 not found}\else
   {\not@eoftrue \chardef\other=12
    \def\do##1{\catcode`##1=\other}\dospecials \catcode`\ =10
    \loop
       \if@psfile
	  \read\ps@stream to \epsf@fileline
       \else{
	  \obeyspaces
          \read\ps@stream to \epsf@tmp\global\let\epsf@fileline\epsf@tmp}
       \fi
       \ifeof\ps@stream\not@eoffalse\else
       \if@psfile\else
       \expandafter\epsf@test\epsf@fileline:. \\%
       \fi
          \expandafter\epsf@aux\epsf@fileline:. \\%
       \fi
   \ifnot@eof\repeat
   }\closein\ps@stream\fi}%
\long\def\epsf@test#1#2#3:#4\\{\def\epsf@testit{#1#2}
			\ifx\epsf@testit\epsf@start\else
\ps@typeout{Warning! File does not start with `\epsf@start'.  It may not be a PostScript file.}
			\fi
			\@psfiletrue} 
{\catcode`\%=12\global\let\epsf@percent=
\long\def\epsf@aux#1#2:#3\\{\ifx#1\epsf@percent
   \def\epsf@testit{#2}\ifx\epsf@testit\epsf@bblit
	\@atendfalse
        \epsf@atend #3 . \\%
	\if@atend	
	   \if@verbose{
		\ps@typeout{psfig: found `(atend)'; continuing search}
	   }\fi
        \else
        \epsf@grab #3 . . . \\%
        \not@eoffalse
        \global\no@bbfalse
        \fi
   \fi\fi}%
\def\epsf@grab #1 #2 #3 #4 #5\\{%
   \global\def\epsf@llx{#1}\ifx\epsf@llx\empty
      \epsf@grab #2 #3 #4 #5 .\\\else
   \global\def\epsf@lly{#2}%
   \global\def\epsf@urx{#3}\global\def\epsf@ury{#4}\fi}%
\def\epsf@atendlit{(atend)} 
\def\epsf@atend #1 #2 #3\\{%
   \def\epsf@tmp{#1}\ifx\epsf@tmp\empty
      \epsf@atend #2 #3 .\\\else
   \ifx\epsf@tmp\epsf@atendlit\@atendtrue\fi\fi}

\chardef\psletter = 11 
\chardef\other = 12

\newif \ifdebug 
\newif\ifc@mpute 
\c@mputetrue 

\let\then = \relax
\def\r@dian{pt }
\let\r@dians = \r@dian
\let\dimensionless@nit = \r@dian
\let\dimensionless@nits = \dimensionless@nit
\def\internal@nit{sp }
\let\internal@nits = \internal@nit
\newif\ifstillc@nverging
\def \Mess@ge #1{\ifdebug \then \message {#1} \fi}

{ 
	\catcode `\@ = \psletter
	\gdef \nodimen {\expandafter \n@dimen \the \dimen}
	\gdef \term #1 #2 #3%
	       {\edef \t@ {\the #1}
		\edef \t@@ {\expandafter \n@dimen \the #2\r@dian}%
		\t@rm {\t@} {\t@@} {#3}%
	       }
	\gdef \t@rm #1 #2 #3%
	       {{%
		\count 0 = 0
		\dimen 0 = 1 \dimensionless@nit
		\dimen 2 = #2\relax
		\Mess@ge {Calculating term #1 of \nodimen 2}%
		\loop
		\ifnum	\count 0 < #1
		\then	\advance \count 0 by 1
			\Mess@ge {Iteration \the \count 0 \space}%
			\Multiply \dimen 0 by {\dimen 2}%
			\Mess@ge {After multiplication, term = \nodimen 0}%
			\Divide \dimen 0 by {\count 0}%
			\Mess@ge {After division, term = \nodimen 0}%
		\repeat
		\Mess@ge {Final value for term #1 of 
				\nodimen 2 \space is \nodimen 0}%
		\xdef \Term {#3 = \nodimen 0 \r@dians}%
		\aftergroup \Term
	       }}
	\catcode `\p = \other
	\catcode `\t = \other
	\gdef \n@dimen #1pt{#1} 
}

\def \Divide #1by #2{\divide #1 by #2} 

\def \Multiply #1by #2
       {{
	\count 0 = #1\relax
	\count 2 = #2\relax
	\count 4 = 65536
	\Mess@ge {Before scaling, count 0 = \the \count 0 \space and
			count 2 = \the \count 2}%
	\ifnum	\count 0 > 32767 
	\then	\divide \count 0 by 4
		\divide \count 4 by 4
	\else	\ifnum	\count 0 < -32767
		\then	\divide \count 0 by 4
			\divide \count 4 by 4
		\else
		\fi
	\fi
	\ifnum	\count 2 > 32767 
	\then	\divide \count 2 by 4
		\divide \count 4 by 4
	\else	\ifnum	\count 2 < -32767
		\then	\divide \count 2 by 4
			\divide \count 4 by 4
		\else
		\fi
	\fi
	\multiply \count 0 by \count 2
	\divide \count 0 by \count 4
	\xdef \product {#1 = \the \count 0 \internal@nits}%
	\aftergroup \product
       }}

\def\r@duce{\ifdim\dimen0 > 90\r@dian \then   
		\multiply\dimen0 by -1
		\advance\dimen0 by 180\r@dian
		\r@duce
	    \else \ifdim\dimen0 < -90\r@dian \then  
		\advance\dimen0 by 360\r@dian
		\r@duce
		\fi
	    \fi}

\def\Sine#1%
       {{%
	\dimen 0 = #1 \r@dian
	\r@duce
	\ifdim\dimen0 = -90\r@dian \then
	   \dimen4 = -1\r@dian
	   \c@mputefalse
	\fi
	\ifdim\dimen0 = 90\r@dian \then
	   \dimen4 = 1\r@dian
	   \c@mputefalse
	\fi
	\ifdim\dimen0 = 0\r@dian \then
	   \dimen4 = 0\r@dian
	   \c@mputefalse
	\fi
	\ifc@mpute \then
		\divide\dimen0 by 180
		\dimen0=3.141592654\dimen0
		\dimen 2 = 3.1415926535897963\r@dian 
		\divide\dimen 2 by 2 
		\Mess@ge {Sin: calculating Sin of \nodimen 0}%
		\count 0 = 1 
		\dimen 2 = 1 \r@dian 
		\dimen 4 = 0 \r@dian 
		\loop
			\ifnum	\dimen 2 = 0 
			\then	\stillc@nvergingfalse 
			\else	\stillc@nvergingtrue
			\fi
			\ifstillc@nverging 
			\then	\term {\count 0} {\dimen 0} {\dimen 2}%
				\advance \count 0 by 2
				\count 2 = \count 0
				\divide \count 2 by 2
				\ifodd	\count 2 
				\then	\advance \dimen 4 by \dimen 2
				\else	\advance \dimen 4 by -\dimen 2
				\fi
		\repeat
	\fi		
			\xdef \sine {\nodimen 4}%
       }}

\def\Cosine#1{\ifx\sine\UnDefined\edef\Savesine{\relax}\else
		             \edef\Savesine{\sine}\fi
	{\dimen0=#1\r@dian\advance\dimen0 by 90\r@dian
	 \Sine{\nodimen 0}
	 \xdef\cosine{\sine}
	 \xdef\sine{\Savesine}}}	      

\def\psdraft{
	\def\@psdraft{0}
}
\def\psfull{
	\def\@psdraft{100}
}

\psfull

\newif\if@scalefirst
\def\psscalefirst{\@scalefirsttrue}
\def\psrotatefirst{\@scalefirstfalse}
\psrotatefirst

\newif\if@draftbox
\def\psnodraftbox{
	\@draftboxfalse
}
\def\psdraftbox{
	\@draftboxtrue
}
\@draftboxtrue

\newif\if@prologfile
\newif\if@postlogfile
\def\pssilent{
	\@noisyfalse
}
\def\psnoisy{
	\@noisytrue
}
\psnoisy
\newif\if@bbllx
\newif\if@bblly
\newif\if@bburx
\newif\if@bbury
\newif\if@height
\newif\if@width
\newif\if@rheight
\newif\if@rwidth
\newif\if@angle
\newif\if@clip
\newif\if@verbose
\def\@p@@sclip#1{\@cliptrue}
\newif\if@decmpr
\def\@p@@sfigure#1{\def\@p@sfile{null}\def\@p@sbbfile{null}\@decmprfalse
   \openin1=\ps@predir#1
   \ifeof1
	\closein1
	\get@dir{#1}
	\ifx\ps@founddir\leer
		\openin1=\ps@predir#1.bb
		\ifeof1
			\closein1
			\get@dir{#1.bb}
			\ifx\ps@founddir\leer
				\ps@typeout{Can't find #1 in \figurepath}
			\else
				\@decmprtrue
				\def\@p@sfile{\ps@founddir\ps@dir#1}
				\def\@p@sbbfile{\ps@founddir\ps@dir#1.bb}
			\fi
		\else
			\closein1
			\@decmprtrue
			\def\@p@sfile{#1}
			\def\@p@sbbfile{#1.bb}
		\fi
	\else
		\def\@p@sfile{\ps@founddir\ps@dir#1}
		\def\@p@sbbfile{\ps@founddir\ps@dir#1}
	\fi
   \else
	\closein1
	\def\@p@sfile{#1}
	\def\@p@sbbfile{#1}
   \fi
}
\def\@p@@sfile#1{\@p@@sfigure{#1}}
\def\@p@@sbbllx#1{
		\@bbllxtrue
		\dimen100=#1
		\edef\@p@sbbllx{\number\dimen100}
}
\def\@p@@sbblly#1{
		\@bbllytrue
		\dimen100=#1
		\edef\@p@sbblly{\number\dimen100}
}
\def\@p@@sbburx#1{
		\@bburxtrue
		\dimen100=#1
		\edef\@p@sbburx{\number\dimen100}
}
\def\@p@@sbbury#1{
		\@bburytrue
		\dimen100=#1
		\edef\@p@sbbury{\number\dimen100}
}
\def\@p@@sheight#1{
		\@heighttrue
		\dimen100=#1
   		\edef\@p@sheight{\number\dimen100}
}
\def\@p@@swidth#1{
		\@widthtrue
		\dimen100=#1
		\edef\@p@swidth{\number\dimen100}
}
\def\@p@@srheight#1{
		\@rheighttrue
		\dimen100=#1
		\edef\@p@srheight{\number\dimen100}
}
\def\@p@@srwidth#1{
		\@rwidthtrue
		\dimen100=#1
		\edef\@p@srwidth{\number\dimen100}
}
\def\@p@@sangle#1{
		\@angletrue
		\edef\@p@sangle{#1} 
}
\def\@p@@ssilent#1{ 
		\@verbosefalse
}
\def\@p@@sprolog#1{\@prologfiletrue\def\@prologfileval{#1}}
\def\@p@@spostlog#1{\@postlogfiletrue\def\@postlogfileval{#1}}
\def\@cs@name#1{\csname #1\endcsname}
\def\@setparms#1=#2,{\@cs@name{@p@@s#1}{#2}}
\def\ps@init@parms{
		\@bbllxfalse \@bbllyfalse
		\@bburxfalse \@bburyfalse
		\@heightfalse \@widthfalse
		\@rheightfalse \@rwidthfalse
		\def\@p@sbbllx{}\def\@p@sbblly{}
		\def\@p@sbburx{}\def\@p@sbbury{}
		\def\@p@sheight{}\def\@p@swidth{}
		\def\@p@srheight{}\def\@p@srwidth{}
		\def\@p@sangle{0}
		\def\@p@sfile{} \def\@p@sbbfile{}
		\def\@p@scost{10}
		\def\@sc{}
		\@prologfilefalse
		\@postlogfilefalse
		\@clipfalse
		\if@noisy
			\@verbosetrue
		\else
			\@verbosefalse
		\fi
}
\def\parse@ps@parms#1{
	 	\@psdo\@psfiga:=#1\do
		   {\expandafter\@setparms\@psfiga,}}
\newif\ifno@bb
\def\bb@missing{
	\if@verbose{
		\ps@typeout{psfig: searching \@p@sbbfile \space  for bounding box}
	}\fi
	\no@bbtrue
	\epsf@getbb{\@p@sbbfile}
        \ifno@bb \else \bb@cull\epsf@llx\epsf@lly\epsf@urx\epsf@ury\fi
}	
\def\bb@cull#1#2#3#4{
	\dimen100=#1 bp\edef\@p@sbbllx{\number\dimen100}
	\dimen100=#2 bp\edef\@p@sbblly{\number\dimen100}
	\dimen100=#3 bp\edef\@p@sbburx{\number\dimen100}
	\dimen100=#4 bp\edef\@p@sbbury{\number\dimen100}
	\no@bbfalse
}
\newdimen\p@intvaluex
\newdimen\p@intvaluey
\def\rotate@#1#2{{\dimen0=#1 sp\dimen1=#2 sp
		  \global\p@intvaluex=\cosine\dimen0
		  \dimen3=\sine\dimen1
		  \global\advance\p@intvaluex by -\dimen3
		  \global\p@intvaluey=\sine\dimen0
		  \dimen3=\cosine\dimen1
		  \global\advance\p@intvaluey by \dimen3
		  }}
\def\compute@bb{
		\no@bbfalse
		\if@bbllx \else \no@bbtrue \fi
		\if@bblly \else \no@bbtrue \fi
		\if@bburx \else \no@bbtrue \fi
		\if@bbury \else \no@bbtrue \fi
		\ifno@bb \bb@missing \fi
		\ifno@bb \ps@typeout{FATAL ERROR: no bb supplied or found}
			\no-bb-error
		\fi
		\count203=\@p@sbburx
		\count204=\@p@sbbury
		\advance\count203 by -\@p@sbbllx
		\advance\count204 by -\@p@sbblly
		\edef\ps@bbw{\number\count203}
		\edef\ps@bbh{\number\count204}
		\if@angle 
			\Sine{\@p@sangle}\Cosine{\@p@sangle}
	        	{\dimen100=\maxdimen\xdef\r@p@sbbllx{\number\dimen100}
					    \xdef\r@p@sbblly{\number\dimen100}
			                    \xdef\r@p@sbburx{-\number\dimen100}
					    \xdef\r@p@sbbury{-\number\dimen100}}
                        \def\minmaxtest{
			   \ifnum\number\p@intvaluex<\r@p@sbbllx
			      \xdef\r@p@sbbllx{\number\p@intvaluex}\fi
			   \ifnum\number\p@intvaluex>\r@p@sbburx
			      \xdef\r@p@sbburx{\number\p@intvaluex}\fi
			   \ifnum\number\p@intvaluey<\r@p@sbblly
			      \xdef\r@p@sbblly{\number\p@intvaluey}\fi
			   \ifnum\number\p@intvaluey>\r@p@sbbury
			      \xdef\r@p@sbbury{\number\p@intvaluey}\fi
			   }
			\rotate@{\@p@sbbllx}{\@p@sbblly}
			\minmaxtest
			\rotate@{\@p@sbbllx}{\@p@sbbury}
			\minmaxtest
			\rotate@{\@p@sbburx}{\@p@sbblly}
			\minmaxtest
			\rotate@{\@p@sbburx}{\@p@sbbury}
			\minmaxtest
			\edef\@p@sbbllx{\r@p@sbbllx}\edef\@p@sbblly{\r@p@sbblly}
			\edef\@p@sbburx{\r@p@sbburx}\edef\@p@sbbury{\r@p@sbbury}
		\fi
		\count203=\@p@sbburx
		\count204=\@p@sbbury
		\advance\count203 by -\@p@sbbllx
		\advance\count204 by -\@p@sbblly
		\edef\@bbw{\number\count203}
		\edef\@bbh{\number\count204}
}
\def\in@hundreds#1#2#3{\count240=#2 \count241=#3
		     \count100=\count240	
		     \divide\count100 by \count241
		     \count101=\count100
		     \multiply\count101 by \count241
		     \advance\count240 by -\count101
		     \multiply\count240 by 10
		     \count101=\count240	
		     \divide\count101 by \count241
		     \count102=\count101
		     \multiply\count102 by \count241
		     \advance\count240 by -\count102
		     \multiply\count240 by 10
		     \count102=\count240	
		     \divide\count102 by \count241
		     \count200=#1\count205=0
		     \count201=\count200
			\multiply\count201 by \count100
		 	\advance\count205 by \count201
		     \count201=\count200
			\divide\count201 by 10
			\multiply\count201 by \count101
			\advance\count205 by \count201
		     \count201=\count200
			\divide\count201 by 100
			\multiply\count201 by \count102
			\advance\count205 by \count201
		     \edef\@result{\number\count205}
}
\def\compute@wfromh{
		\in@hundreds{\@p@sheight}{\@bbw}{\@bbh}
		\edef\@p@swidth{\@result}
}
\def\compute@hfromw{
	        \in@hundreds{\@p@swidth}{\@bbh}{\@bbw}
		\edef\@p@sheight{\@result}
}
\def\compute@handw{
		\if@height 
			\if@width
			\else
				\compute@wfromh
			\fi
		\else 
			\if@width
				\compute@hfromw
			\else
				\edef\@p@sheight{\@bbh}
				\edef\@p@swidth{\@bbw}
			\fi
		\fi
}
\def\compute@resv{
		\if@rheight \else \edef\@p@srheight{\@p@sheight} \fi
		\if@rwidth \else \edef\@p@srwidth{\@p@swidth} \fi
}
\def\compute@sizes{
	\compute@bb
	\if@scalefirst\if@angle
	\if@width
	   \in@hundreds{\@p@swidth}{\@bbw}{\ps@bbw}
	   \edef\@p@swidth{\@result}
	\fi
	\if@height
	   \in@hundreds{\@p@sheight}{\@bbh}{\ps@bbh}
	   \edef\@p@sheight{\@result}
	\fi
	\fi\fi
	\compute@handw
	\compute@resv}
\def\OzTeXSpecials{
	\special{empty.ps /@isp {true} def}
	\special{empty.ps \@p@swidth \space \@p@sheight \space
			\@p@sbbllx \space \@p@sbblly \space
			\@p@sbburx \space \@p@sbbury \space
			startTexFig \space }
	\if@clip{
		\if@verbose{
			\ps@typeout{(clip)}
		}\fi
		\special{empty.ps doclip \space }
	}\fi
	\if@angle{
		\if@verbose{
			\ps@typeout{(rotate)}
		}\fi
		\special {empty.ps \@p@sangle \space rotate \space} 
	}\fi
	\if@prologfile
	    \special{\@prologfileval \space } \fi
	\if@decmpr{
		\if@verbose{
			\ps@typeout{psfig: Compression not available
			in OzTeX version \space }
		}\fi
	}\else{
		\if@verbose{
			\ps@typeout{psfig: including \@p@sfile \space }
		}\fi
		\special{epsf=\ps@predir\@p@sfile \space }
	}\fi
	\if@postlogfile
	    \special{\@postlogfileval \space } \fi
	\special{empty.ps /@isp {false} def}
}
\def\DvipsSpecials{
	\special{ps::[begin] 	\@p@swidth \space \@p@sheight \space
			\@p@sbbllx \space \@p@sbblly \space
			\@p@sbburx \space \@p@sbbury \space
			startTexFig \space }
	\if@clip{
		\if@verbose{
			\ps@typeout{(clip)}
		}\fi
		\special{ps:: doclip \space }
	}\fi
	\if@angle
		\if@verbose{
			\ps@typeout{(clip)}
		}\fi
		\special {ps:: \@p@sangle \space rotate \space} 
	\fi
	\if@prologfile
	    \special{ps: plotfile \@prologfileval \space } \fi
	\if@decmpr{
		\if@verbose{
			\ps@typeout{psfig: including \@p@sfile.Z \space }
		}\fi
		\special{ps: plotfile "`zcat \@p@sfile.Z" \space }
	}\else{
		\if@verbose{
			\ps@typeout{psfig: including \@p@sfile \space }
		}\fi
		\special{ps: plotfile \@p@sfile \space }
	}\fi
	\if@postlogfile
	    \special{ps: plotfile \@postlogfileval \space } \fi
	\special{ps::[end] endTexFig \space }
}
\def\psfig#1{\vbox {
	\ps@init@parms
	\parse@ps@parms{#1}
	\compute@sizes
	\ifnum\@p@scost<\@psdraft{
		\PsfigSpecials 
		\vbox to \@p@srheight sp{
			\hbox to \@p@srwidth sp{
				\hss
			}
		\vss
		}
	}\else{
		\if@draftbox{		
			\hbox{\fbox{\vbox to \@p@srheight sp{
			\vss
			\hbox to \@p@srwidth sp{ \hss 
			 \hss }
			\vss
			}}}
		}\else{
			\vbox to \@p@srheight sp{
			\vss
			\hbox to \@p@srwidth sp{\hss}
			\vss
			}
		}\fi

	}\fi
}}
\psfigRestoreAt
\setDriver
\let\@=\LaTeXAtSign

\documentstyle{cupconf}
\setcounter{page}{1}

\newcommand\etal{{ et al. }}
\def\mdot{$\dot M$}
\def\msy{$M_\odot$ yr$^{-1}$}
\def\kms{km s$^{-1}$}
\def\e#1{$\times$ $10^{#1}$ }
\def\ee#1{$10^{#1}$ }
\def\lsim{\mathrel{\rlap{\lower 4pt \hbox{\hskip 1pt $\sim$}}\raise 1pt \hbox
        {$<$}}}
\def\gsim{\mathrel{\rlap{\lower 4pt \hbox{\hskip 1pt $\sim$}}\raise 1pt \hbox
        {$>$}}}
\def\ltsim{\mathrel{\rlap{\lower 4pt \hbox{\hskip 1pt $\sim$}}\raise 1pt \hbox
        {$<$}}}
\def\gtsim{\mathrel{\rlap{\lower 4pt \hbox{\hskip 1pt $\sim$}}\raise 1pt \hbox
        {$>$}}}
\def\apj{ApJ}
\def\aap{A\&A}
\def\aj{AJ}
\def\mnras{MNRAS}
\def\araa{ARAA}
\def\pasp{PASP}
\def\apjs{ApJS}
\def\nat{Nature}

\title[SN Ia]{Type Ia Supernova Progenitors, Environmental Effects, and\\ 
Cosmic Supernova Rates}

\author[Nomoto, Umeda, Hachisu, Kato, Kobayashi, Tsujimoto]
{Ken'ichi NOMOTO$^1$, Hideyuki UMEDA$^1$, Izumi HACHISU$^2$, Mariko
KATO$^3$, Chiaki KOBAYASHI$^1$, \& Takuji TSUJIMOTO$^4$}
\affiliation{$^1$Department of Astronomy, and Research Center for the 
Early Universe, University of Tokyo \\Tokyo 113-0033 
\\$^2$Department of Earth Science and Astronomy, 
College of Arts and Sciences, University of Tokyo,
Tokyo 153-8902 
\\$^3$Department of Astronomy, Keio University, 
Hiyoshi, Kouhoku-ku, Yokohama 223-8521 
\\$^4$National Astronomical Observatory, Mitaka, Tokyo 181-8588
}

\begin{document}

\maketitle

  \vspace*{-8cm}
  \noindent
  {\small
	To appear in ``Type Ia Supernovae: Theory and Cosmology'' \\
        ed. J. Truran \& J. Niemeyer (Cambridge University Press), 1999}
  \vspace{7.2cm}

\begin{abstract}

Relatively uniform light curves and spectral evolution of Type Ia
supernovae (SNe Ia) have led to the use of SNe Ia as a ``standard
candle'' to determine cosmological parameters, such as the Hubble
constant, the density parameter, and the cosmological constant.
Whether a statistically significant value of the cosmological constant
can be obtained depends on whether the peak luminosities of SNe Ia are
sufficiently free from the effects of cosmic and galactic evolutions.

Here we first review the single degenerate scenario for the
Chandrasekhar mass white dwarf (WD) models of SNe Ia.  We identify the
progenitor's evolution and population with two channels: (1) the WD+RG
(red-giant) and (2) the WD+MS (near main-sequence He-rich star)
channels.  In these channels, the strong wind from accreting white
dwarfs plays a key role, which yields important age and metallicity
effects on the evolution.

We then address the questions whether the nature of SNe Ia depends
systematically on environmental properties such as metallicity and age
of the progenitor system and whether significant evolutionary effects
exist.  We suggest that the variation of the carbon mass fraction
$X$(C) in the C+O WD (or the variation of the initial WD mass) causes
the diversity of the brightness of SNe Ia.  This model can explain 
the observed dependence of SNe Ia brighness on the galaxy types.

Finally, applying the metallicity effect on the evolution of SN Ia
progenitors, we make a prediction of the cosmic supernova rate history
as a composite of the supernova rates in different types of galaxies.

\end{abstract}

\section{Introduction}

Type Ia supernovae (SNe Ia) are good distance indicators, and provide
a promising tool for determining cosmological parameters (e.g., Branch
1998).  SNe Ia have been discovered up to $z \sim 1.32$ (Gilliland et
al. 1999).  Both the Supernova Cosmology Project (Perlmutter \etal
1997, 1999) and the High-z Supernova Search Team (\cite{gar98}; Riess
\etal 1998) have suggested a statistically significant value for the
cosmological constant.

However, SNe Ia are not perfect standard candles, but show some
intrinsic variations in brightness.  When determining the absolute
peak luminosity of high-redshift SNe Ia, therefore, these analyses
have taken advantage of the empirical relation existing between the
peak brightness and the light curve shape (LCS).  Since this relation
has been obtained from nearby SNe Ia only (Phillips 1993; Hamuy \etal
1995; Riess \etal 1995), it is important to examine whether it depends
systematically on environmental properties such as metallicity and age
of the progenitor system.

High-redshift supernovae present us very useful information, not only
to determine cosmological parameters but also to put constraints on
the star formation history in the universe.  They have given the SN Ia
rate at $z \sim 0.5$ (Pain 1999) but will provide the SN Ia rate
history over $0<z<1$.  With the Next Generation Space Telescope, both
SNe Ia and SNe II will be observed through $z \sim 4$.  It is useful
to provide a prediction of cosmic supernova rates to constrain the age
and metallicity effects of the SN Ia progenitors.

SNe Ia have been widely believed to be a thermonuclear explosion of a
mass-accreting white dwarf (WD) (e.g., Nomoto et al. 1997a for a
review).  However, the immediate progenitor binary systems have not
been clearly identified yet (\cite{bra95}).  In order to address the
above questions regarding the nature of high-redshift SNe Ia, we need
to identify the progenitors systems and examine the ``evolutionary''
effects (or environmental effects) on those systems.

In \S2, we summarize the progenitors' evolution where the strong wind
from accreting WDs plays a key role (Hachisu, Kato, \& Nomoto 1996,
1999, hereafter HKN96, HKN99; Hachisu, Kato, Nomoto \& Umeda 1999,
HKNU99).  In \S3, we addresses the issue of whether a difference in
the environmental properties is at the basis of the observed range of
peak brightness (Umeda et al. 1999b).  In \S4, we make a prediction of
the cosmic supernova rate history as a composite of the different
types of galaxies (Kobayashi et al. 1999).

\section{Progenitor's evolution and the white dwarf wind}

There exist two models proposed as progenitors of SNe Ia: 1) the
Chandrasekhar mass model, in which a mass-accreting carbon-oxygen
(C+O) WD grows in mass up to the critical mass $M_{\rm Ia} \simeq
1.37-1.38 M_\odot$ near the Chandrasekhar mass and explodes as an SN
Ia (e.g., Nomoto \etal 1984, 1994), and 2) the sub-Chandrasekhar mass
model, in which an accreted layer of helium atop a C+O WD ignites
off-center for a WD mass well below the Chandrasekhar mass (e.g.,
Arnett 1996). The early time spectra of the majority of SNe Ia are in
excellent agreement with the synthetic spectra of the Chandrasekhar
mass models, while the spectra of the sub-Chandrasekhar mass models
are too blue to be comparable with the observations (\cite{hof96};
\cite{nug97}).

     For the evolution of accreting WDs toward the Chandrasekhar mass,
two scenarios have been proposed: 1) a double degenerate (DD)
scenario, i.e., merging of double C+O WDs with a combined mass
surpassing the Chandrasekhar mass limit (\cite{ibe84}; \cite{web84}),
and 2) a single degenerate (SD) scenario, i.e., accretion of
hydrogen-rich matter via mass transfer from a binary companion (e.g.,
\cite{nom82}; Nomoto et al. 1994).  The issue of DD vs. SD is still 
debated (e.g., \cite{bra95}), although theoretical modeling has
indicated that the merging of WDs leads to the accretion-induced
collapse rather than SN Ia explosion (Saio \& Nomoto 1985, 1998;
\cite{seg97}).

     In the SD Chandrasekhar mass model for SNe Ia, a WD explodes as a
SN Ia only when its rate of the mass accretion ($\dot M$) is in a
certain narrow range (e.g., Nomoto 1982; Nomoto \& Kondo 1991).  In
particular, if $\dot M$ exceeds the critical rate $\dot M_{\rm b}$ in
Eq.(2.1) below, the accreted matter extends to form a common envelope
(Nomoto et al. 1979).  This difficulty has been overcome by the WD
wind model (see below).  For the actual binary systems which grow the
WD mass ($M_{\rm WD}$) to $M_{\rm Ia}$, the following two systems are
appropriate.  One is a system consisting of a mass-accreting WD and a
lobe-filling, more massive, slightly evolved main-sequence (MS) or
sub-giant star (hereafter ``WD+MS system'').  The other system
consists of a WD and a lobe-filling, less massive, red-giant
(hereafter ``WD+RG system'').

\subsection {White dwarf winds}

\begin{figure}
\centerline{\psfig{figure=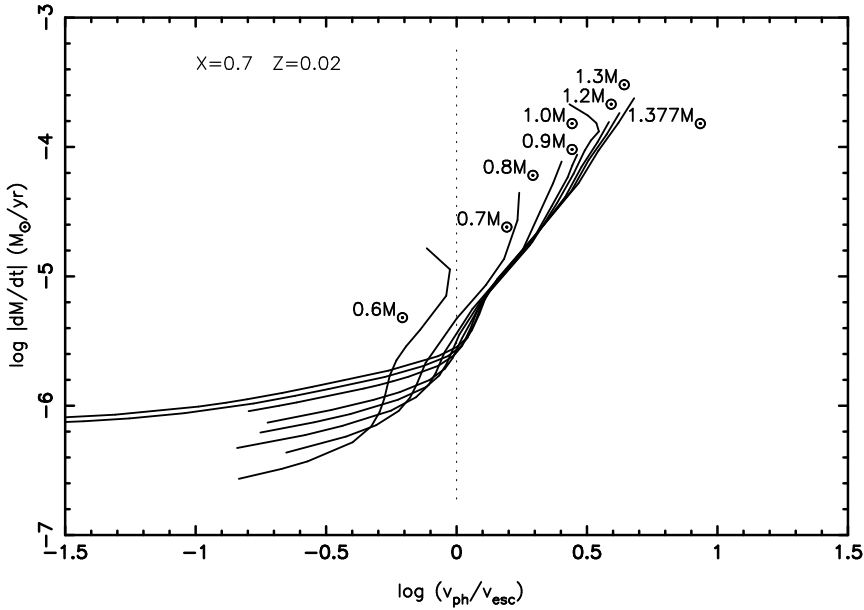,width=12cm}}
\caption[h]{\label{dmdtescx70z02}
Ratio of the photospheric velocity to the escape 
velocity there $v_{\rm ph} / v_{\rm esc}$ 
is plotted against the decreasing rate of 
the envelope mass for WDs with masses of 
$0.6 M_\odot$, $0.7 M_\odot$, $0.8 M_\odot$, $0.9 M_\odot$, 
$1.0 M_\odot$, $1.2 M_\odot$, $1.3 M_\odot$, and 
$1.377 M_\odot$.   
We regard the wind as ``strong'' when the photospheric 
velocity exceeds the escape velocity there, i.e.,
$v_{\rm ph} > v_{\rm esc}$.  If not, it is regarded as
``weak.'' 
}
\end{figure}

Optically thick WD winds are driven when the accretion rate $\dot M$
exceeds the critical rate $\dot M_{\rm b}$.  Here $\dot M_{\rm b}$ is
the rate at which steady burning can process the accreted hydrogen
into helium as
\begin{equation}
\dot M_{\rm b} \approx 0.75 \times10^{-6} \left({M_{\rm WD} \over {M_\odot}}
 - 0.40\right) M_\odot {\rm ~yr}^{-1}.
\label{critical_rate}
\end{equation}

With such a rapid accretion, the WD envelope expands to $R_{\rm ph}
\sim 0.1 R_\odot$ and the photospheric temperature decreases below
$\log T_{\rm ph} \sim 5.5$.  Around this temperature, the shoulder of
the strong peak of OPAL Fe opacity (\cite{igl93}) drives the
radiation-driven wind (HKN96; HKN99).  We plot the ratio $v_{\rm ph}/
v_{\rm esc}$ between the photospheric velocity and the escape velocity
at the photosphere in Figure \ref{dmdtescx70z02} against the mass
transfer rate.  We call the wind {\sl strong} when $v_{\rm ph} >
v_{\rm esc}$. When the wind is strong, $v_{\rm ph} \sim 1000$ km
s$^{-1}$ being much faster than the orbital velocity.

If the wind is sufficiently strong, the WD can avoid the formation of
a common envelope and steady hydrogen burning increases its mass
continuously at a rate $\dot M_{\rm b}$ by blowing the extra mass away
in a wind.  When the mass transfer rate decreases below this critical
value, optically thick winds stop.  If the mass transfer rate further
decreases below $\sim$ 0.5 $\dot M_{\rm b}$, hydrogen shell burning
becomes unstable to trigger very weak shell flashes but still burns
a large fraction of accreted hydrogen.

     The steady hydrogen shell burning converts hydrogen into helium
atop the C+O core and increases the mass of the helium layer
gradually.  When its mass reaches a certain value, weak helium shell
flashes occur.  Then a part of the envelope mass is blown off but a
large fraction of He can be burned to C+O (Kato \& Hachisu 1999) to
increase the WD mass.  In this way, Thus strong winds from the
accreting WD play a key role to increase the WD mass to $M_{\rm Ia}$.

\begin{figure}
\centerline{\psfig{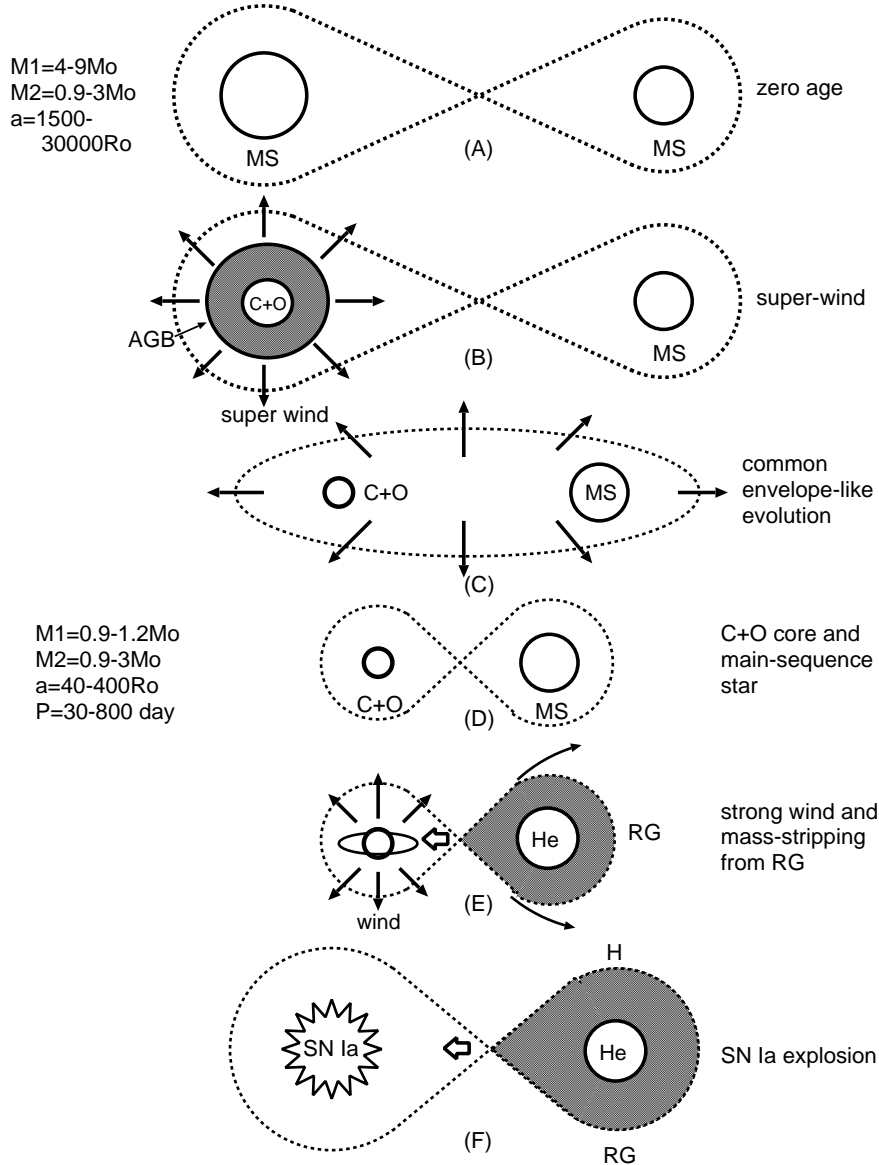}}
\caption{
An illustration of the WD+RG (symbiotic) channel to Type Ia supernovae.
\label{channel}}
\end{figure}

\subsection{WD+RG system} 

This is a symbiotic binary system consisting of a white dwarf (WD) and
a low mass red-giant (RG).  A full evolutionary path of the WD+RG
system from the zero age main-sequence stage ({\it stage A}) to the SN
Ia explosion ({\it stage F}) is as follows.  

\begin{enumerate}
\item[(A)] Zero age main-sequence.
\item[(B)] The primary has evolved first to become an asymptotic giant 
branch (AGB) star and blows a slow wind (or a super wind) at the end
of the AGB evolution.
\item[(C)] If the superwind from the AGB star is as fast as 
or slower than the orbital velocity, the wind outflowing from the
system takes away the orbital angular momentum effectively.  As a
result the wide binary shrinks greatly (by about a factor of ten or
more) to become a close binary.  This is a similar process to the
common envelope evolution.
\item[(D)] Then the AGB star undergoes a common envelope 
evolution.  The AGB star forms a C+O WD, and the initial secondary
remains a main-sequence star (MS).
\item[(E)] The initial secondary evolves to a red-giant (RG) forming 
a helium core and fills up its inner critical Roche lobe.  Mass
transfer begins.  The WD component blows a strong wind and the winds
can stabilize the mass transfer even if the RG component has a deep
convective envelope.
\item[(F)] The WD component has grown in mass to $M_{\rm Ia}$
and explodes as a Type Ia supernova.
\end{enumerate}

     The occurrence frequency of SNe Ia through this channel is much
larger than the earlier scenario, because of the following two
evolutionary processes, which have not considered before.

(1) Because of the AGB wind at stage C, the WD + RG close binary can
form from a wide binary even with such a large initial separation as
$a_i \lsim 40,000 R_\odot$.  Our earlier estimate (HKN96) is
constrained by $a_i \lsim 1,500 R_\odot$.

(2) When the RG fills its inner critical Roche lobe, the WD undergoes
rapid mass accretion and blows a strong optically thick wind.  Our
earlier analysis has shown that the mass transfer is stabilized by
this wind only when the mass ratio of RG/WD is smaller than 1.15.  Our
new finding is that the WD wind can strip mass from the RG envelope,
which could be efficient enough to stabilize the mass transfer even if
the RG/WD mass ratio exceeds 1.15.  If this mass-stripping effect is
strong enough, though its efficiency $\eta_{\rm eff}$ is subject to
uncertainties, the symbiotic channel can produce SNe Ia for a much
(ten times or more) wider range of the binary parameters than our
earlier estimation.

With the above two new effects (1) and (2), the WD+RG (symbiotic)
channel can account for the inferred rate of SNe Ia in our Galaxy.
The immediate progenitor binaries in this symbiotic channel to SNe Ia
may be observed as symbiotic stars, luminous supersoft X-ray sources,
or recurrent novae like T CrB or RS Oph, depending on the wind status.

\subsection{WD+MS system} 

\begin{figure}
\centerline{\psfig{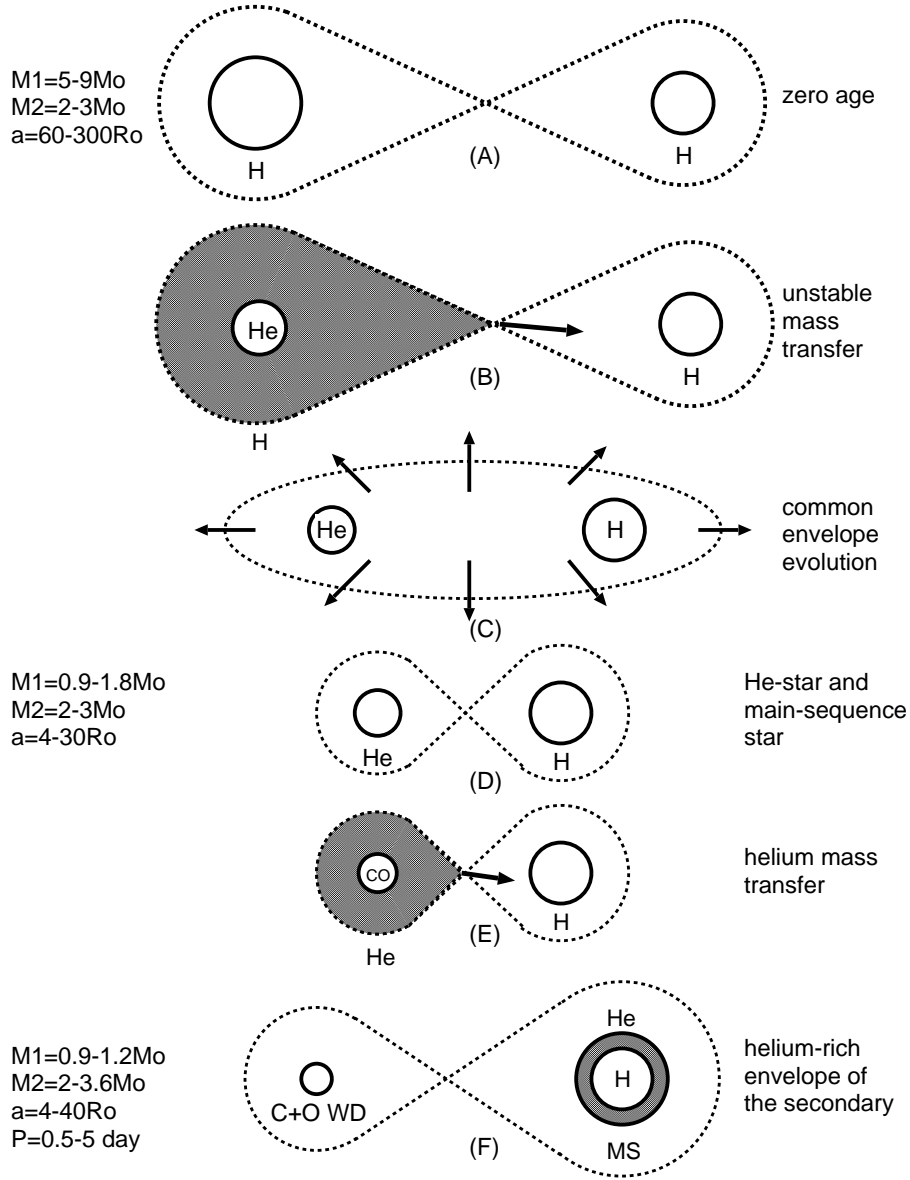}}
\caption{
An illustration of the WD+MS channel to Type Ia supernovae
through the common envelope evolution to the helium matter transfer.
\label{hemtsn1a}}
\end{figure}

In this scenario, a C+O WD is originated, not from an AGB star with a
C+O core, but from a red-giant star with a helium core of $\sim
0.8-2.0 M_\odot$.  The helium star, which is formed after the first
common envelope evolution, evolves to form a C+O WD of $\sim 0.8-1.1
M_\odot$ with transferring a part of the helium envelope onto the
secondary main-sequence star.

As an example for this system, let us consider a pair of $7 M_\odot +
2.5 M_\odot$ with the initial separation of $a_i \sim 50-600 R_\odot$.
The binary evolves to SN Ia through the following stages
(stages A-F in Figure \ref{hemtsn1a} and G in HKNU99):
\begin{enumerate}
\item[(1) stage A-C:]  
When the mass of the helium core grows to $1.0 M_\odot < M_{\rm 1, He}
< 1.4 M_\odot$, the primary fills its Roche lobe and the binary
undergoes a common envelope evolution.
\item[(2) stage C-D:] 
After the common envelope evolution, the system consists of a helium
star and a main-sequence star with a relatively compact separation of
$a_f \sim 3-40 ~R_\odot$ and $P_{\rm orb} \sim 0.4 - 20$ d.
\item[(3) stage D:]
The helium star contracts and ignites central helium burning to become
a helium main-sequence star.  The primary stays at the helium
main-sequence for $\sim 1 \times 10^7$ yr.
\item[(4) stage E:]
After helium exhaustion, a carbon-oxygen core develops.  When the core
mass reaches $0.9-1.0 M_\odot$, the helium star evolves to a red-giant
and fills again its inner critical Roche lobe.  Almost pure helium is
transferred to the secondary because the mass transfer is stable for
the mass ratio $q=M_1/M_2 < 0.79$.  The mass transfer rate is as large
as $\sim 1 \times 10^{-5} M_\odot$ yr$^{-1}$ but the mass-receiving
main-sequence star ($\sim 2-3 M_\odot$) does not expand for such a low
rate.
\item[(5) stage F:]
The secondary has received $0.1-0.4 M_\odot$ (almost) pure helium and,
as a result, it becomes a helium-rich star as observed in the
recurrent nova U Sco.  The primary becomes a C+O WD.  The separation
and thus the orbital period gradually increases during the mass
transfer phase.  The final orbital period becomes $P_{\rm orb}
\sim 0.5-40$ d.
\item[(6) stage G:]
The white dwarf accretes hydrogen-rich, helium-enhanced matter from a
lobe-filling, slightly evolved companion at a critical rate and blows
excess matter in the wind. The white dwarf grows in mass to $M_{\rm
Ia}$ and explodes as an SN Ia.

\end{enumerate}

This evolutionary path provides a much wider channel to SNe Ia than
previous scenarios.  A part of the progenitor systems are identified
as the luminous supersoft X-ray sources (van den Heuvel et al. 1992)
during steady H-burning (but without wind to avoid extinction), or the
recurrent novae like U Sco if H-burning is weakly unstable.  Actually
these objects are characterized by the accretion of helium-rich
matter.

\begin{figure}
\centerline{\psfig{figure=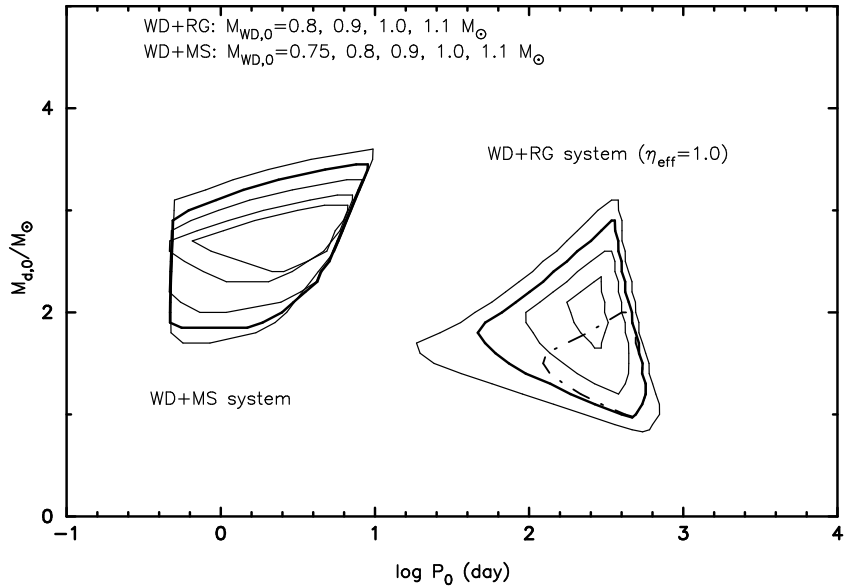,width=12cm}}
\caption[h]{\label{ztotreg100}
The region to produce SNe Ia in the $\log P_0 - M_{\rm d,0}$ plane for
five initial white dwarf masses of $0.75 M_\odot$, $0.8 M_\odot$, $0.9
M_\odot$, $1.0 M_\odot$ (heavy solid line), and $1.1 M_\odot$.  The
region of $M_{\rm WD,0}= 0.7 M_\odot$ almost vanishes for both the
WD+MS and WD+RG systems, and the region of $M_{\rm WD,0}= 0.75
M_\odot$ vanishes for the WD+RG system.  Here, we assume the stripping
efficiency of $\eta_{\rm eff}=1$.  For comparison, we show only the
region of $M_{\rm WD,0}= 1.0 M_\odot$ for a much lower efficiency of
$\eta_{\rm eff}=0.3$ by a dash-dotted line.
}
\end{figure}

\subsection{Realization Frequency}

\begin{figure}
\centerline{\psfig{figure=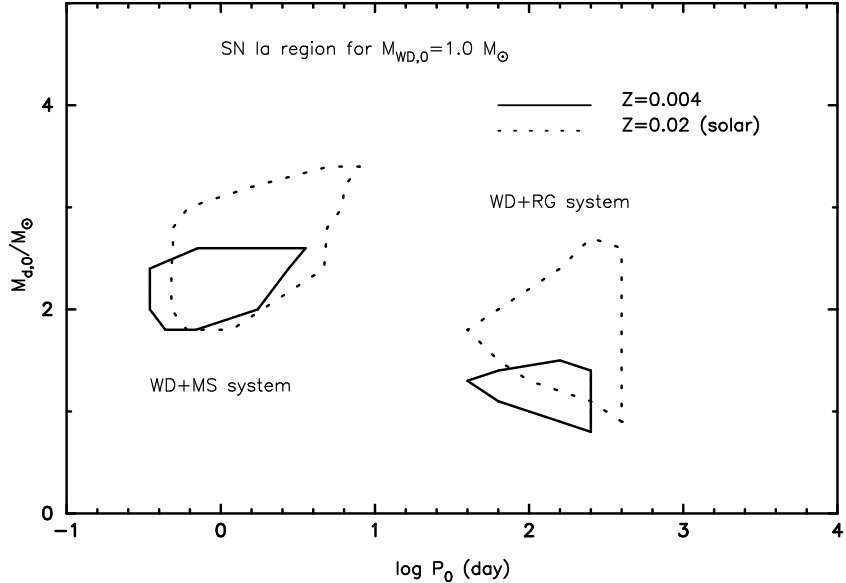,width=12cm}}
\caption[h]{\label{koba_fig2}
The regions of SNe Ia is plotted in the initial orbital period vs. the
initial companion mass diagram for the initial WD mass of $M_{\rm
WD,0}=1.0 M_\odot$.  The dashed and solid lines represent the cases of
solar abundance ($Z=0.02$) and much lower metallicity of $Z=0.004$,
respectively. The left and the right regions correspond to the WD+MS
and the WD+RG systems, respectively.
}
\end{figure}

\begin{figure}
\centerline{\psfig{figure=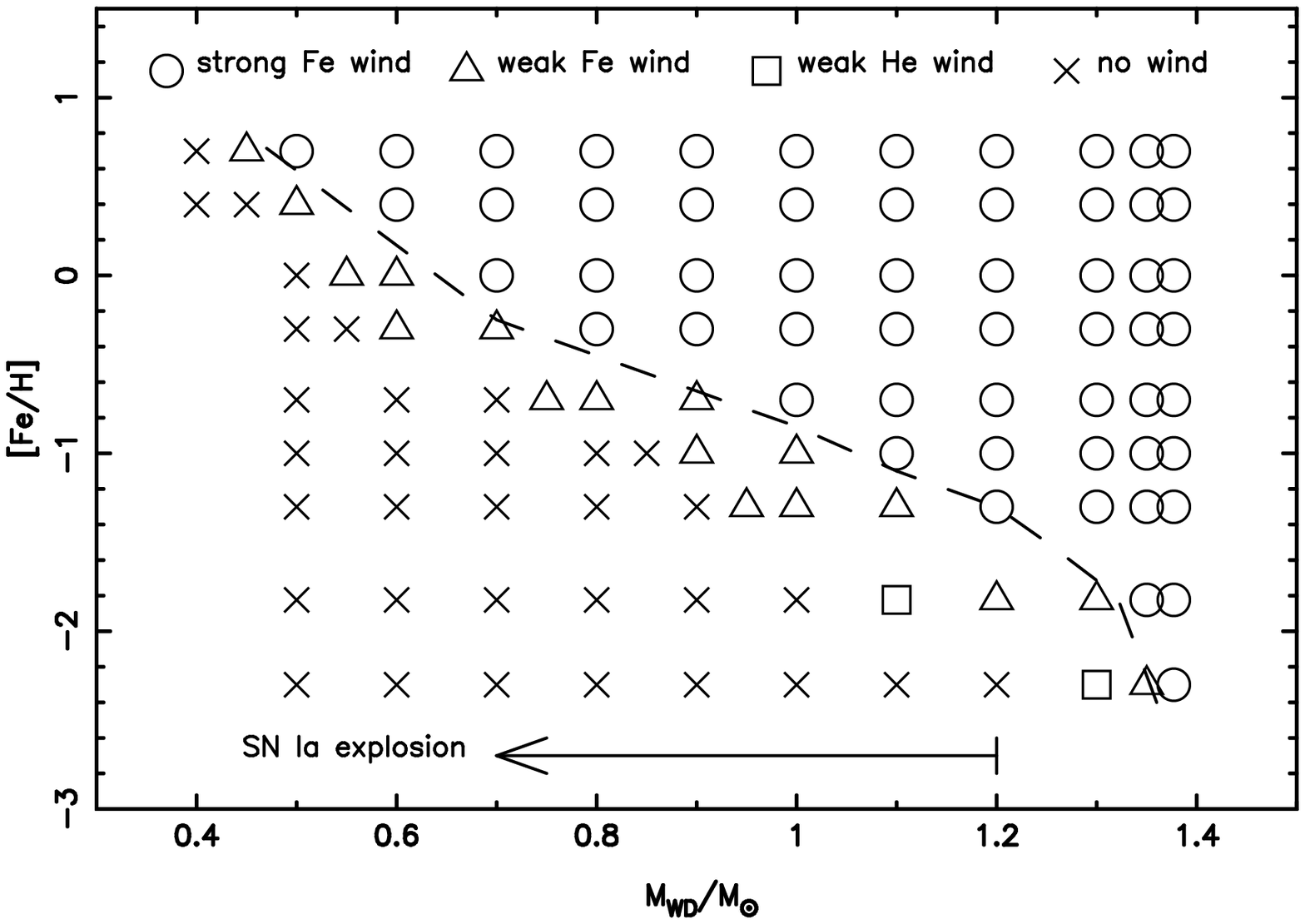,width=12cm}}
\caption[h]{\label{koba_fig1}
WD mass vs. metallicity diagram showing the metallicity dependence of
optically thick winds We regard the wind as ``strong'' if the wind
velocity at the photosphere exceeds the escape velocity but ``weak''
if the wind velocity is lower than the escape velocity.  The term of
``He'' or ``Fe'' wind denotes that the wind is accelerated by the peak
of iron lines near $\log T ({\rm K}) \sim 5.2$ or of helium lines near
$\log T ({\rm K}) \sim 4.6$.  The dashed line indicates the
demarcation between the ``strong'' wind and the ``weak'' wind.  }
\end{figure}

     For an immediate progenitor system WD+RG of SNe Ia, we consider a
close binary initially consisting of a C+O WD with $M_{\rm WD,0}=
0.6-1.2 M_{\odot}$ and a low-mass red-giant star with $M_{\rm RG,0}=
0.7-3.0 M_{\odot}$ having a helium core of $M_{\rm He,0}= 0.2-0.46
M_{\odot}$ (stage E). The initial state of these immediate progenitors
is specified by three parameters, i.e., $M_{\rm WD,0}$, $M_{\rm RG,0}
= M_{\rm d,0}$, and the initial orbital period $P_0$ ($M_{\rm He,0}$
is determined if $P_0$ is given).

     We follow binary evolutions of these systems and obtain the
parameter range(s) which can produce an SN Ia.  In Figure
\ref{ztotreg100}, the region enclosed by the thin solid line produces
SNe Ia for several cases of the initial WD mass, $M_{\rm WD,0} =$ 0.75
- 1.1 $M_\odot$.  For smaller $M_{\rm WD,0}$, the wind is weaker, so
that the SN Ia region is smaller. The regions of $M_{\rm WD,0} = 0.6
M_\odot$ and $0.7 M_\odot$ vanish for both the WD+MS and WD+RG
systems.  

     In the outside of this region, the outcome of the evolution at
the end of the calculations is not an SN Ia but one of the followings:
\begin{enumerate}
\item[(i)]
  Formation of a common envelope for too large $M_{\rm d}$ or $P_0
\sim$ day, where the mass transfer is unstable at the beginning of
mass transfer.
\item[(ii)] 
  Novae or strong hydrogen shell flash for too small $M_{\rm d,0}$,
where the mass transfer rate becomes below \ee{-7} \msy.
\item[(iii)]
  Helium core flash of the red giant component for too long $P_0$,
where a central helium core flash ignites, i.e., the helium core mass
of the red-giant reaches $0.46 M_\odot$.
\item[(iv)]
  Accretion-induced collapse for $M_{\rm WD,0} > 1.2 M_\odot$, where
the central density of the WD reaches $\sim 10^{10}$ g cm$^{-3}$
before heating wave from the hydrogen burning layer reaches the
center.  As a result, the WD undergoes collapse due to electron
capture without exploding as an SN Ia (\cite{nom91}).
\end{enumerate}

     It is clear that the new region of the WD+RG system is not
limited by the condition of $q < 1.15$, thus being ten times or more
wider than the region of HKN96's model (depending on the the stripping
efficiency of $\eta_{\rm eff}$).  

     The WD+MS progenitor system can also be specified by three
initial parameters: the initial C+O WD mass $M_{\rm WD,0}$, the mass
donor's initial mass $M_{\rm d,0}$, and the orbital period $P_0$.  For
$M_{\rm WD,0} = 1.0 M_\odot$, the region producing an SN Ia is bounded
by $M_{\rm d,0}= 1.8-3.2 M_\odot$ and $P_0= 0.5-5$ d as shown by the
solid line in Figure \ref{ztotreg100}.  The upper and lower bounds are
respectively determined by the common envelope formation (i) and
nova-like explosions (ii) as above.  The left and right bounds are
determined by the minimum and maximum radii during the main sequence
of the donor star (HKNU99).

     We estimate the rate of SNe Ia originating from these channels in
our Galaxy by using equation (1) of Iben \& Tutukov (1984).  The
realization frequencies of SNe Ia through the WD+RG and WD+MS channels
are estimated as $\sim$ 0.0017 yr$^{-1}$ (WD+RG) and $\sim$ 0.001
yr$^{-1}$ (WD+MS), respectively.  The total SN Ia rate of the
WD+MS/WD+RG systems becomes $\sim$ 0.003 yr$^{-1}$, which is close
enough to the inferred rate of our Galaxy.

\subsection{Low metallicity inhibition of type Ia supernovae}

In the above SN Ia progenitor model, the accreting WD blows a strong
wind to reach the Chandrasekhar mass limit.  If the iron abundance of
the progenitors is as low as [Fe/H]$\lsim - 1$, then the wind is too
weak for SNe Ia to occur.

     In this model, an interesting metallicity effect has been found
(Kobayashi \etal 1998; Hachisu \& Kato 1999).  The wind velocity is
higher for larger $M_{\rm WD}$ and larger Fe/H because of higher
luminosity and larger opacity, respectively.  In order to blow
sufficiently strong wind ($v_{\rm w} > v_{\rm esc}$), $M_{\rm WD}$
should exceed a certain mass $M_{\rm w}$ (Fig. \ref{dmdtescx70z02}).
As seen from the dashed line in Figure \ref{koba_fig1}, $M_{\rm w}$ is
larger for lower metallicity; e.g., $M_{\rm w} =$ 0.65, 0.85, and 0.95
$M_\odot$ for $Z =$ 0.02, 0.01, and 0.004, respectively. In order for
a WD to grow its mass at $\dot M > \dot M_{\rm b}$, its initial mass
$M_{\rm WD,0}$ should exceed $M_{\rm w}$.  In other words, $M_{\rm w}$
is the metallicity-dependent minimum $M_{\rm WD,0}$ required for a WD
to become an SN Ia.  

\subsection{Possible detection of hydrogen}

In our scenario, the WD winds form a circumstellar envelope around the
binary systems prior to the explosion, which may emit X-rays, radio,
and H$\alpha$ lines by shock heating when the ejecta collide with the
circumstellar envelope. The mass accretion rate in the present models
is still as high as 1 \e{-6} \msy~ for some of the white dwarfs near
the Chandrasekhar limit, so that such a white dwarf explode in the
strong wind phase.

     Our strong wind model of case P1 predicts the presence of
circumstellar matter around the exploding white dwarf.  Whether such a
circumstellar matter is observable depends on its density.  The wind
mass loss rate from the white dwarf near the Chandrasekhar limit is as
high as \mdot~ $\sim$ 1 \e{-8}$-$1 \e{-7} \msy~ and the wind velocity
is $v=$ 1000 \kms.  Despite the relatively high mass loss rate, the
circumstellar density is not so high because of the high wind
velocity.  For steady wind, the density is expressed by $\dot M /v$
($=4 \pi r^2 \rho$).  Normalized by the typical red-giant wind
velocity of 10 \kms, the density measure of our white dwarf wind is
given as \mdot$/v_{10} \sim 1$ \e{-10}$-1 $\e{-9}
\msy, where $v_{10} = v$/10 \kms.

Behind the red-giant, matter stripped from the red-giant component
forms a much dense circumstellar tail.  Its rate is as large as $\sim
1 \times 10^{-7} M_\odot$ yr$^{-1}$ with the velocity of $\sim 100$ km
s$^{-1}$.  The density measure of the dense red-giant wind thus formed
is \mdot/$v_{10} \sim$ 1 \e{-8} \msy.

Further out, circumstellar matter is produced from the wind from the
red-giant companion, which is too far away to cause significant
circumstellar interaction.

For some cases, winds from the WD have stopped before the explosion.
Therefore, circumstellar matter is dominated by the wind from the
red-giant companion whose velocity is as low as $\sim 10$ km s$^{-1}$.

At SN Ia explosion, ejecta would collide with the circumstellar
matter, which produces shock waves propagating both outward and
inward.  At the shock front, particle accelerations take place to
cause radio emissions.  Hot plasmas in the shocked materials emit
thermal X-rays.  The circumstellar matter ahead of the shock is
ionized by X-rays and produce recombination H$\alpha$ emissions
(\cite{cum96}).  Such an interaction has been observed in Type Ib, Ic,
and II supernovae, most typically in SN 1993J (e.g., \cite{suz95} and
references therein).

For SNe Ia, several attempts have been made to detect the above
signature of circumstellar matter.  There has been no radio and X-ray
detections so far.  The upper limit set by X-ray observations of SN
1992A is \mdot$/v_{10} = (2-3)$ \e{-6} \msy~ (\cite{sch93}).  Radio
observations of SN 1986G have provided the most strict upper limit to
the circumstellar density as \mdot/$v_{10} =$ 1 \e{-7} \msy~
(Eck et al. 1995).  This is still $10-100$ times higher than the density
predicted for the white dwarf wind.  If the WD wind has ceased and the
wind mass loss rate from the red-giant is significantly higher than 1
\e{-7} \msy, radio detection could be possible for very nearby SNe Ia
as close as SN 1986G.  (Note also that SN 1986G is not a typical SN Ia
but a subluminous SN Ia.)

For H$\alpha$ emissions, Branch et al. (1983) noted a small, narrow
emission feature at the rest wavelength of H$\alpha$, which is
blueshifted by 1800 \kms~ from the local interstellar Ca II
absorption.  Though this feature was not observed 5 days later, such
high velocity hydrogen is expected from the white dwarf wind model.
For SN 1994D, Cumming et al. (1996) obtained the upper limit of
\mdot/$v_{10} =$ 6 \e{-6} \msy.  Further attempts to detect H$\alpha$
emissions are highly encouraged.

\section{The origin of diversity of SNe Ia and environmental effects}

 There are some observational indications that SNe Ia are affected by
their environment. The most luminous SNe Ia seem to occur only in
spiral galaxies, while both spiral and elliptical galaxies are hosts
for dimmer SNe Ia. Thus the mean peak brightness is dimmer in
ellipticals than in spiral galaxies (Hamuy \etal 1996). The SNe Ia
rate per unit luminosity at the present epoch is almost twice as high
in spirals as in ellipticals (Cappellaro \etal 1997).  Moreover, Wang
\etal (1997) and Riess \etal (1999) found that the variation of the
peak brightness for SNe located in the outer regions in galaxies is
smaller.

H\"oflich \etal (1998, 1999) examined how the initial composition of
the WD (metallicity and the C/O ratio) affects the observed properties
of SNe Ia.  Umeda \etal (1999a) obtained the C/O ratio as a function
of the main-sequence mass and metallicity of the WD progenitors.
Umeda et al. (1999b) suggested that the variation of the C/O ratio is
the main cause of the variation of SNe Ia brightness, with larger C/O
ratio yielding brighter SNe Ia.  We will show that the C/O ratio
depends indeed on environmental properties, such as the metallicity
and age of the companion of the WD, and that our model can explain
most of the observational trends discussed above. We then make some
predictions about the brightness of SN Ia at higher redshift.

\subsection{C/O ratio in WD progenitors}

\begin{figure} 
\centerline{\psfig{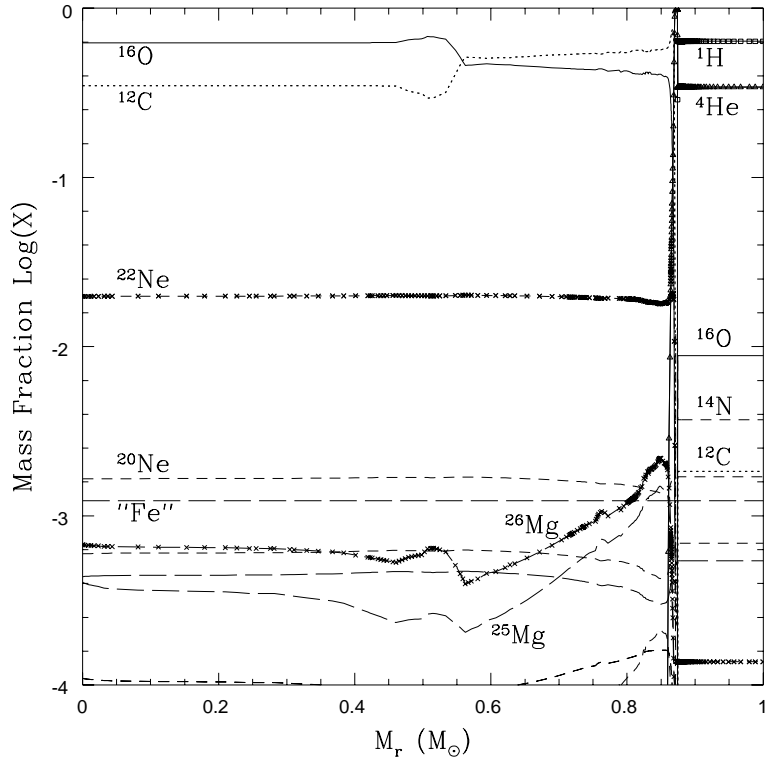}}
\caption[h]{
Abundances in mass fraction in the inner core of the 6 $M_\odot$ star
for $Y$ = 0.2775 and $Z$ = 0.02 at the end of the second dredge-up.  
\label{umefig1}}
\end{figure}

 In this section we discuss how the C/O ratio in the WD depends on the
metallicity and age of the binary system.  The C/O ratio in C+O WDs
depends primarily on the main-sequence mass of the WD progenitor and
on metallicity. 

 We calculated the evolution of intermediate-mass ($3-9 M_\odot$)
stars for metallicity $Z$=0.001 -- 0.03.  In the ranges of stellar
masses and $Z$ considered in this paper, the most important
metallicity effect is that the radiative opacity is smaller for lower
$Z$. Therefore, a star with lower $Z$ is brighter, thus having a
shorter lifetime than a star with the same mass but higher $Z$.  In
this sense, the effect of reducing metallicity for these stars is
almost equivalent to increasing a stellar mass.

 For stars with larger masses and/or smaller $Z$, the luminosity is
higher at the same evolutionary phase.  With a higher nuclear energy
generation rate, these stars have larger convective cores during H and
He burning, thus forming larger He and C-O cores.

 As seen in Figure \ref{umefig1}, the central part of these stars
is oxygen-rich. The C/O ratio is nearly constant in the innermost
region, which was a convective core during He burning.  Outside this
homogeneous region, where the C-O layer grows due to He shell burning,
the C/O ratio increases up to C/O $\gsim 1$; thus the oxygen-rich core
is surrounded by a shell with C/O $\gsim$ 1. In fact this is a generic
feature in all models we calculated. The C/O ratio in the shell is C/O
$\simeq$ 1 for the star as massive as $\sim 7 M_\odot$, and C/O $>1$
for less massive stars.

\begin{figure} 
\centerline{\psfig{figure=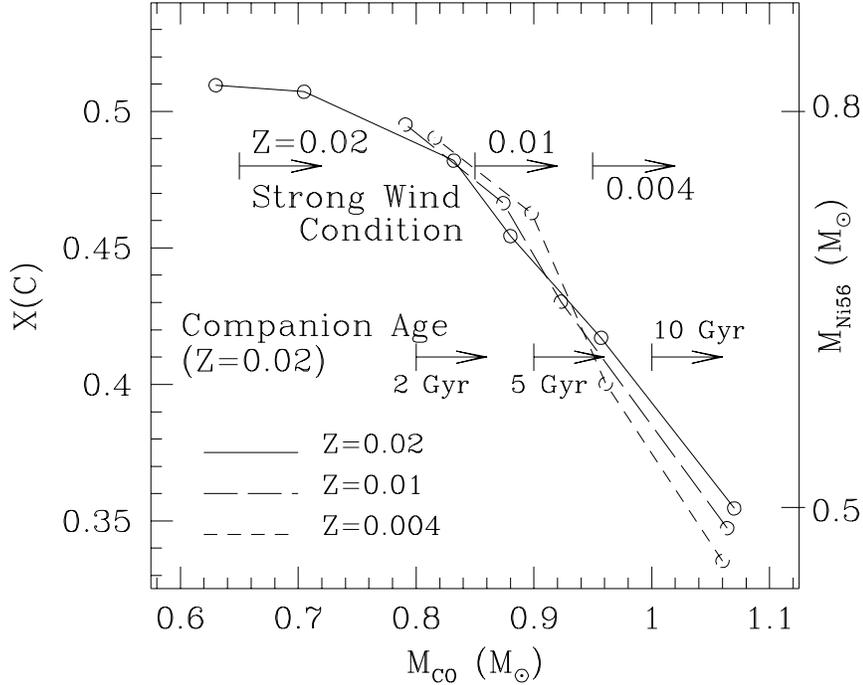,width=12cm}}
\caption[h]{
The total $^{12}$C mass fraction included in the convective core of
mass, $M=1.14M_\odot$, just before the SN Ia explosion as a function
of the C+O core mass before the onset of mass accretion, 
$M_{\rm CO}$. The lower bounds of $M_{\rm CO}$ obtained from the age
effects and the conditions for strong wind to blow are also shown by
arrows.
\label{umefig2}}
\end{figure}

When a progenitor reaches the critical mass for the SNe Ia explosion,
the central core is convective up to around 1.1 $M_\odot$. Hence the
relevant C/O ratio is between the central value before convective
mixing and the total C/O of the whole WD. Using the results from the
C6 model (Nomoto \etal 1984), we assume that the convective region is
1.14 $M_\odot$ and for simplicity, C/O = 1 outside the C-O core at the
end of second dredge-up. Then we obtain the C/O ratio of the inner
part of the SNe Ia progenitors (Fig. \ref{umefig2}).

 From this figure we find three interesting trends.  First, while the
central C/O is a complicated function of stellar mass (Umeda \etal
1999a), as shown here the C/O ratio in the core before SNe Ia explosion
is a decreasing monotonic function of mass.  The central C/O ratio at
the end of second dredge-up decreases with mass for $M_{\rm ms} \gsim
5M_\odot$, while the ratio increases with mass for $M_{\rm ms} \gsim
4M_\odot$; however, the convective core mass during He burning is
smaller for a less massive star, and the C/O ratio during shell He
burning is larger for smaller C+O core. Hence, when the C/O ratio is
averaged over 1.1 $M_\odot$ the C/O ratio decreases with mass. Second,
as shown in Umeda \etal 1999, although the C/O ratio is a complicated
function of metallicity and mass, the metallicity dependence is
remarkably converged when the ratio is seen as a function of the C+O
core mass ($M_{\rm CO}$) instead of the initial main sequence mass.

According to the evolutionary calculations for 3$-$9 $M_\odot$ stars
by Umeda \etal (1999a), the C/O ratio and its distribution are
determined in the following evolutionary stages of the close binary.

(1) At the end of central He burning in the 3$-$9 $M_\odot$ primary
star, C/O$<1$ in the convective core. The mass of the core is larger
for more massive stars. 

(2) After central He exhaustion, the outer C+O layer grows via He
shell burning, where C/O$\gsim 1$ (Umeda \etal 1999a).

(3a) If the primary star becomes a red giant (case C evolution;
e.g. van den Heuvel 1994), it then undergoes the second dredge-up,
forming a thin He layer, and enters the AGB phase. The C+O core mass,
$M_{\rm CO}$, at this phase is larger for more massive stars. For a
larger $M_{\rm CO}$ the total carbon mass fraction is smaller. 

(3b) When it enters the AGB phase, the star greatly expands and is
assumed here to undergo Roche lobe overflow (or a super-wind phase)
and to form a C+O WD. Thus the initial mass of the WD, $M_{\rm
WD,0}$, in the close binary at the beginning of mass accretion is
approximately equal to $M_{\rm CO}$.

(4a) If the primary star becomes a He star (case BB evolution), the
second dredge-up in (3a) corresponds to the expansion of the He
envelope.

(4b) The ensuing Roche lobe overflow again leads to a white dwarf of
mass $M_{\rm WD,0}$ = $M_{\rm CO}$.

(5) After the onset of mass accretion, the WD mass grows through
steady H burning and weak He shell flashes, as described in the WD
wind model.  The composition of the growing C+O layer is assumed to be
C/O=1.

(6) The WD grows in mass and ignites carbon when its mass reaches
$M_{\rm Ia} =1.367 M_\odot$, as in the model C6 of Nomoto \etal
(1984).  Because of strong electron-degeneracy, carbon burning is
unstable and grows into a deflagration for a central temperature of
$8\times 10^8$ K and a central density of $1.47\times 10^9$ g
cm$^{-3}$.  At this stage, the convective core extends to $M_r =
1.14M_\odot$ and the material is mixed almost uniformly, as in the C6
model.

In Figure \ref{umefig2}, we show the carbon mass fraction $X$(C) in
the convective core of this pre-explosive WD, as a function of
metallicity ($Z$) and initial mass of the WD before the onset of mass
accretion, $M_{\rm CO}$.  Figure \ref{umefig2} reveals that: 1) $X$(C)
is smaller for larger $M_{\rm CO} \simeq M_{\rm WD,0}$.  2) The
dependence of $X$(C) on metallicity is small when plotted against
$M_{\rm CO}$, even though the relation between $M_{\rm CO}$ and the
initial stellar mass depends sensitively on $Z$ (Umeda \etal 1999a).

\subsection{Brightness of SNe Ia and the C/O ratio}

In the Chandrasekhar mass models for SNe Ia, the brightness of SNe Ia
is determined mainly by the mass of $^{56}$Ni synthesized ($M_{\rm
Ni56}$). Observational data suggest that $M_{\rm Ni56}$ for most SNe
Ia lies in the range $M_{\rm Ni56} \sim 0.4 - 0.8 M_\odot$
(e.g. Mazzali \etal 1998). This range of $M_{\rm Ni56}$ can result
from differences in the C/O ratio in the progenitor WD as follows.

 In the deflagration model, a larger C/O ratio leads to the production
of more nuclear energy and buoyancy force, thus leading to a faster
propagation.  The faster propagation of the convective deflagration
wave results in a larger $M_{\rm Ni56}$. For example, a variation of
the propagation speed by 15\% in the W6 -- W8 models results in
$M_{\rm Ni56}$ values ranging between 0.5 and $0.7 M_\odot$ (Nomoto
\etal 1984), which could explain the observations.

In the delayed detonation model, $M_{\rm Ni56}$ is predominantly
determined by the deflagration-to-detonation-transition (DDT) density
$\rho_{\rm DDT}$, at which the initially subsonic deflagration turns
into a supersonic detonation (Khokhlov 1991).  As discussed in Umeda
et al. (1999b), $\rho_{\rm DDT}$ could be very sensitive to $X$(C),
and a larger $X$(C) is likely to result in a larger $\rho_{\rm DDT}$
and $M_{\rm Ni56}$.

Here we postulate that $M_{\rm Ni56}$ and consequently brightness of a
SN Ia increase as the progenitors' C/O ratio increases (and thus
$M_{\rm WD,0}$ decreases).  As illustrated in Figure \ref{umefig2},
the range of $M_{\rm Ni56} \sim 0.5-0.8 M_\odot$ is the result of an
$X$(C) range $0.35-0.5$, which is the range of $X$(C) values of our
progenitor models.  The $X$(C) -- $M_{\rm Ni56}$ -- $M_{\rm WD,0}$
relation we adopt is still only a working hypothesis, which needs to
be proved from studies of the turbulent flame during explosion
(e.g., Niemeyer \& Hillebrandt 1995).

\subsection{Metallicity and age effects}

\begin{figure}
\centerline{\psfig{figure=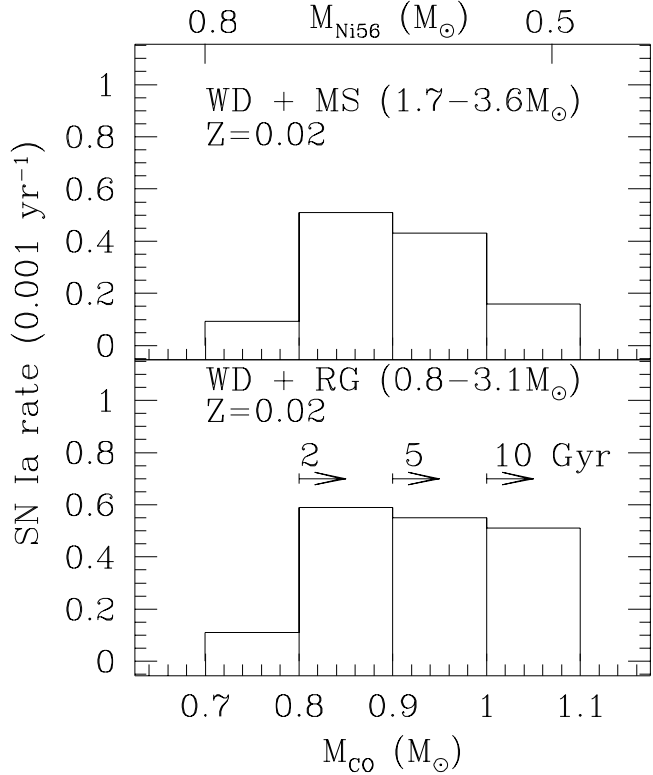,width=10cm}}
\caption[h]{SN Ia frequency for a galaxy of mass $2 \times 10^{11} M_\odot$
as a function of $M_{\rm CO}$ for $Z$=0.02.
For the WD+RG system, constraints from the companion's age are
shown by the arrows. SNe Ia from the WD+MS system occur in spirals
but not in ellipticals because of the age effect. $M_{\rm CO}$ and
$M_{\rm Ni56}$ is assumed to be related as shown here.
\label{umefig3}}
\end{figure}

\subsubsection{Metallicity effects on the minimum $M_{\rm WD,0}$}

As mentioned in \S2.5, $M_{\rm w}$ is the metallicity-dependent
minimum $M_{\rm WD,0}$ for a WD to become an SN Ia ({\sl strong wind
condition} in Fig. \ref{umefig2}).  The upper bound $M_{\rm WD,0}
\simeq 1.07M_\odot$ is imposed by the condition that carbon should not
ignite and is almost independent of metallicity.  As shown in Figure
\ref{umefig2}, the range of $M_{\rm CO} \simeq M_{\rm WD,0}$ can be
converted into a range of $X$(C). From this we find the following
metallicity dependence for $X$(C):

(1) The upper bound of $X$(C), which is determined by the lower limit
on $M_{\rm CO}$ imposed by the metallicity-dependent conditions for a
strong wind, e.g., $X$(C) $\lsim 0.51$, 0.46 and 0.41, for $Z$=0.02,
0.01, and 0.004, respectively.

(2) On the other hand, the lower bound, $X$(C) $\simeq 0.35-0.33$,
does not depend much on $Z$, since it is imposed by the maximum
$M_{\rm CO}$.

(3) Assuming the relation between $M_{\rm Ni56}$ and $ X$(C) given in
Figure \ref{umefig2}, our model predicts the absence of brighter SNe
Ia in lower metallicity environment.

\subsubsection{Age effects on the minimum $M_{\rm WD,0}$} 

In our model, the age of the progenitor system also constrains the
range of $X$(C) in SNe Ia. In the SD scenario, the lifetime of the
binary system is essentially the main-sequence lifetime of the
companion star, which depends on its initial mass $M_2$. HKNU99 and
HKN99 have obtained a constraint on $M_2$ by calculating the evolution
of accreting WDs for a set of initial masses of the WD ($M_{\rm WD,0}
\simeq M_{\rm CO}$) and of the companion ($M_2$), and the initial
binary period ($P_0$). In order for the WD mass to reach $M_{\rm Ia}$,
the donor star should transfer enough material at the appropriate
accretion rates.  The donors of successful cases are divided into two
categories: one is composed of slightly evolved main-sequence stars
with $M_2 \sim 1.7 - 3.6M_\odot$ (for $Z$=0.02), and the other of
red-giant stars with $M_2 \sim 0.8 - 3.1M_\odot$ (for $Z$=0.02)
(Fig. \ref{ztotreg100}).

If the progenitor system is older than 2 Gyr, it should be a system
with a donor star of $M_2 < 1.7 M_\odot$ in the red-giant branch.
Systems with $M_2 > 1.7 M_\odot$ become SNe Ia in a time shorter than
2 Gyr.  Likewise, for a given age of the progenitor system, $M_2$ must
be smaller than a limiting mass. This constraint on $M_2$ can be
translated into the presence of a minimum $M_{\rm CO}$ for a given
age, as follows: For a smaller $M_2$, i.e. for the older system, the
total mass which can be transferred from the donor to the WD is
smaller. In order for $M_{\rm WD}$ to reach $M_{\rm Ia}$, therefore,
the initial mass of the WD, $M_{\rm WD,0} \simeq M_{\rm CO}$, should
be larger.  This implies that the older system should have larger
minimum $M_{\rm CO}$ as indicated in Figure \ref{umefig2}.  Using the
$X$(C)-$M_{\rm CO}$ and $M_{\rm Ni56}$-$X$(C) relations
(Fig. \ref{umefig2}), we conclude that WDs in older progenitor systems
have a smaller $X$(C), and thus produce dimmer SNe Ia.

\subsection{Comparison with observations}

 The first observational indication which can be compared with our
model is the possible dependence of the SN brightness on the
morphology of the host galaxies.  Hamuy \etal (1996) found that the
most luminous SNe Ia occur in spiral galaxies, while both spiral and
elliptical galaxies are hosts to dimmer SNe Ia. Hence, the mean peak
brightness is lower in elliptical than in spiral galaxies.

 In our model, this property is simply understood as the effect of the
different age of the companion. In spiral galaxies, star formation
occurs continuously up to the present time. Hence, both WD+MS and
WD+RG systems can produce SNe Ia. In elliptical galaxies, on the other
hand, star formation has long ended, typically more than 10 Gyr
ago. Hence, WD+MS systems can no longer produce SNe Ia. In Figure
\ref{umefig3}, we show the frequency of the expected SN I for a galaxy
of mass $2 \times 10^{11} M_\odot$ for WD+MS and WD+RG systems
separately as a function of $M_{\rm CO}$. Here we use the results of
HKN99 and HKNU99, and the $M_{\rm CO}-X$(C) and $M_{\rm Ni56}- X$(C)
relations given in Figure \ref{umefig2}.  Since a WD with smaller
$M_{\rm CO}$ is assumed to produce a brighter SN Ia (larger $M_{\rm Ni
56}$), our model predicts that dimmer SNe Ia occur both in spirals and
in ellipticals, while brighter ones occur only in spirals.  The mean
brightness is smaller for ellipticals and the total SN Ia rate per
unit luminosity is larger in spirals than in ellipticals.  These
properties are consistent with observations.

 The second observational suggestion is the radial distribution of SNe
Ia in galaxies. Wang \etal (1997) and Riess \etal (1998) found that
the variation of the peak brightness for SNe Ia located in the outer
regions in galaxies is smaller. This behavior can be understood as the
effect of metallicity.  As shown in Figure \ref{umefig2}, even when
the progenitor age is the same, the minimum $M_{\rm CO}$ is larger for
a smaller metallicity because of the metallicity dependence of the WD
winds. Therefore, our model predicts that the maximum brightness of
SNe Ia decreases as metallicity decreases.  Since the outer regions of
galaxies are thought to have lower metallicities than the inner
regions (Zaritsky et al. 1994; Kobayashi \& Arimoto 1999), our model
is consistent with observations. Wang \etal (1997) also claimed that
SNe Ia may be deficient in the bulges of spiral galaxies. This can be
explained by the age effect, because the bulge consists of old
population stars.

\subsection{Diversity of high redshift supernovae}

 We have suggested that $X$(C) is the quantity very likely to cause
the diversity in $M_{\rm Ni56}$ and thus in the brightness of SNe Ia.
We have then shown that our model predicts that the brightness of SNe
Ia depends on the environment, in a way which is qualitatively
consistent with the observations. Further studies of the propagation
of the turbulent flame and the DDT are necessary in order to actually
prove that $X$(C) is the key parameter.

 Our model predicts that when the progenitors belong to an old
population, or to a low metal environment, the number of very bright
SNe Ia is small, so that the variation in brightness is also smaller.
In spiral galaxies, the metallicity is significantly smaller at
redshifts $z\gsim 1$, and thus both the mean brightness of SNe Ia and
its range tend to be smaller.  At $z\gsim 2$ SNe Ia would not occur in
spirals at all because the metallicity is too low.  In elliptical
galaxies, on the other hand, the metallicity at redshifts $z \sim 1-3$
is not very different from the present value.  However, the age of the
galaxies at $z\simeq 1$ is only about 5 Gyr, so that the mean
brightness of SNe Ia and its range tend to be larger at $z\gsim 1$
than in the present ellipticals because of the age effect.

 We note that the variation of $X$(C) is larger in metal-rich nearby
spirals than in high redshift galaxies.  Therefore, if $X$(C) is the
main parameter responsible for the diversity of SNe Ia, and if the
light curve shape (LCS) method is confirmed by the nearby SNe Ia data,
the LCS method can also be used to determine the absolute magnitude of
high redshift SNe Ia.

\subsection{Possible evolutionary effects}

In the above subsections, we consider the metallicity effects only on
the C/O ratio; this is just to shift the main-sequence mass - $M_{\rm
WD,0}$ relation, thus resulting in no important evolutionary effect.
However, some other metallicity effects could give rise to 
evolution of SNe Ia between high and low redshifts (i.e., between
low and high metallicities).

Here we point out just one possible metallicity effect on the carbon
ignition density in the accreting WD.  The ignition density is
determined by the competition between the compressional heating due to
accretion and the neutrino cooling.  The neutrino emission is enhanced
by the {\sl local} Urca shell process of, e.g., $^{21}$Ne--$^{21}$F
pair (Paczy\'nski 1973).  (Note that this is different from the {\sl
convective} Urca neutrino process).  For higher metallicity, the
abundance of $^{21}$Ne is larger so that the cooling is larger.  This
could delay the carbon ignition until a higher central density is
reached (Nomoto et al. 1997d).

Since the WD with a higher central density has a larger binding
energy, the kinetic energy of SNe Ia tends to be smaller if the same
amount of $^{56}$Ni is produced.  This might cause a systematically
slower light curve evolution at higher metallicity environment.  The
carbon ignition process including these metallicity effects as well as
the convective Urca neutrino process need to be studied (see also
Iwamoto et al. 1999 for nucleosynthesis constraints on the ignition
density).

\section{The chemical evolution in the solar neighborhood}

\begin{figure}
\centerline{\psfig{figure=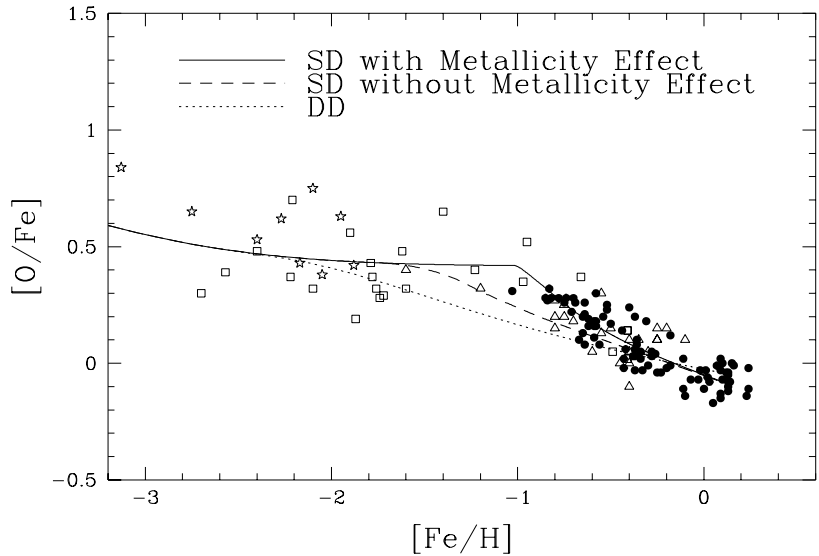,width=12cm}}
\caption[h]{\label{fig:sn}
The evolutionary change in [O/Fe] against [Fe/H] for three SN Ia models. 
The dotted line is for the DD scenario, and
the other lines are for our SD scenario with (solid line) and 
without (dashed line) the metallicity effect on SNe Ia.
Observational data sources: 
filled circles, \cite{edv93};
open triangles, \cite{bar89};
stars, \cite{nis94};
open squares, \cite{gra91}.
}
\end{figure}

The role of SNe II and SNe Ia in the chemical evolution of galaxies
can be seen in the [O/Fe]-[Fe/H] relation (Fig. \ref{fig:sn}:
Metal-poor stars with [Fe/H] $\ltsim -1$ have [O/Fe] $\sim 0.45$ on
the average, while disk stars with [Fe/H] $\gtsim -1$ show a decrease
in [O/Fe] with increasing metallicity.  To explain such an
evolutionary change in [O/Fe] against [Fe/H], we use the chemical
evolution model that allows the infall of material from outside the
disk region. The infall rate, the SFR, and the initial mass function
(IMF) are given by Kobayashi et al. (1998), and the nucleosynthesis
yields of SNe Ia and II are taken from Nomoto et al. (1997bc) and
Tsujimoto et al. (1995).

The metallicity effects on SNe Ia on the chemical evolution is
examined.  For the DD scenario, the distribution function of the
lifetime of SNe Ia by Tutukov \& Yungelson (1994) is adopted, majority
of which is $\sim 0.1-0.3$ Gyr.  Figure \ref{fig:sn} shows the
evolutionary change in [O/Fe] for three SN Ia models.  The dotted line
is for the DD scenario.  The other lines are for our SD scenario with
(solid line) and without (dashed line) the metallicity effect on SNe
Ia.

\begin{enumerate}
\item[(i)] In the DD scenario the lifetime of the majority of SNe Ia 
is shorter than $0.3$ Gyr.  Then the decrease in [O/Fe] starts at
[Fe/H] $\sim -2$, which is too early compared with the observed
decrease in [O/Fe] starting at [Fe/H] $\sim -1$.

\item[(ii)] For the SD scenario with no metallicity effect, the
companion star with $M \sim 2.6 M_\odot$ evolves off the main-sequence
to give rise to SNe Ia at the age of $\sim 0.6$ Gyr.  The resultant
decrease in [O/Fe] starts too early to be compatible with the
observations. 

\item[(iii)] For the metallicity dependent SD scenario, SNe Ia
occur at [Fe/H] $\gtsim -1$, which naturally reproduce the observed
break in [O/Fe] at [Fe/H] $\sim -1$.  Also the low-metallicity
inhibition of SN Ia provides a new interpretation of the SN II-like
abundance patterns of the Galactic halo and the DLA systems
(\cite{kob98}).
\end{enumerate}

\section{Cosmic supernova rates}

Attempts have been made to predict the cosmic supernova rates as a
function of redshift by using the observed cosmic star formation rate
(SFR) (\cite{rui98}; \cite{sad98}; \cite{yun98}).  The observed cosmic
SFR shows a peak at $z \sim 1.4$ and a sharp decrease to the present
(Madau et al. 1996). However, UV luminosities which is converted to
the SFRs may be affected by the dust extinction (\cite{pet98}).
Recent updates of the cosmic SFR suggest that a peak lies around $z
\sim 3$.

Kobayashi et al. (1998) predicts that the cosmic SN Ia rate drops at
$z \sim 1-2$, due to the metallicity-dependence of the SN Ia rate.  
Their finding that the occurrence of SNe Ia depends on the metallicity
of the progenitor systems implies that the SN Ia rate strongly depends
on the history of the star formation and metal-enrichment.
The universe is composed of different morphological types of galaxies
and therefore the cosmic SFR is a sum of the SFRs for different types
of galaxies.  As each morphological type has a unique star formation
history, we should decompose the cosmic SFR into the SFR belonging to
each type of galaxy and calculate the SN Ia rate for each type of
galaxy.

Here we first construct the detailed evolution models for different
type of galaxies which are compatible with the stringent observational
constraints, and apply them to reproduce the cosmic SFR for two
different environments, e.g., the cluster and the field.  Secondly
with the self-consistent galaxy models, we calculate the SN rate
history for each type of galaxy and predict the cosmic supernova rates
as a function of redshift.
We adopt $H_0=50$ km s$^{-1}$ Mpc$^{-1}$, $\Omega_0=0.2$, $\lambda_0=0$
and the galactic age of $15$ Gyr for a standard model, which corresponds 
to the redshift at the formation epoch of galaxies $z_{\rm f}\sim 5$.

\subsection{Supernova rates and galaxy types}

\begin{figure}
\centerline{\psfig{figure=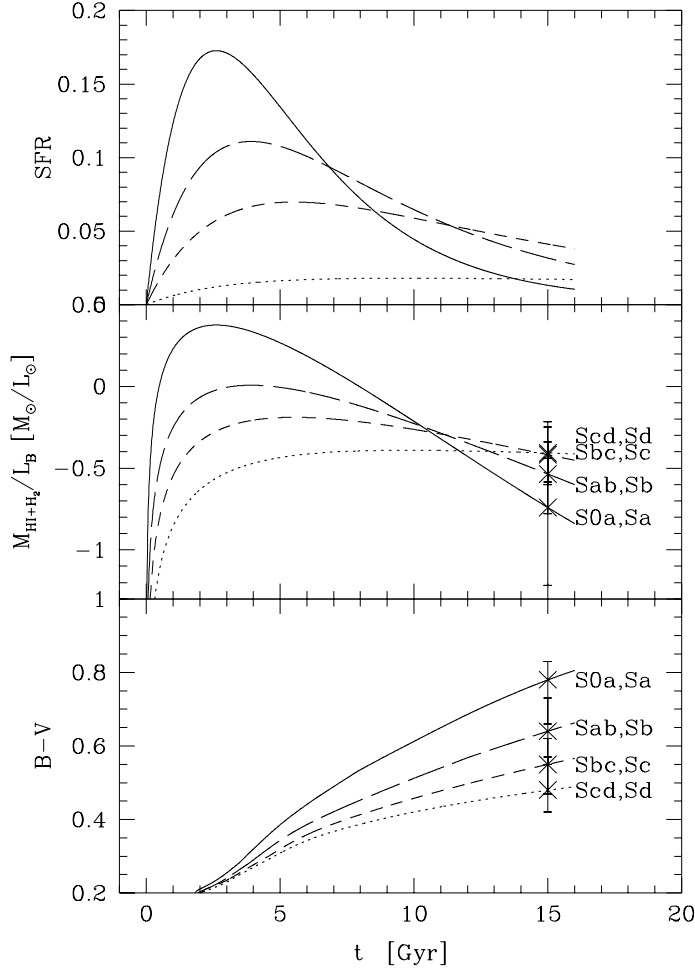,width=14cm}}
\caption[fig2.ps]{\label{fig:sgascol}
The star formation rates (top panel) gas fractions (middle panel) and
the $B-V$ colors (bottom panel) in four types of spirals : S0a-Sa
(solid line), Sab-Sb (long-dashes line), Sbc-Sc (short-dashed line),
and Scd-Sd (dotted line).  The present B-V colors are taken from
Roberts \& Haynes (1994), and the present gas fractions are normalized
by the present blue luminosities and are derived from HI fractions
(\cite{rob94}) and H$_2$/HI ratios (\cite{cas98}).}
\end{figure}

We assume that elliptical galaxies are formed by a single star burst
and stop the star formation at $t \sim 1$ Gyr due to the
supernova-driven galactic wind (e.g., Kodama \& Arimoto 1997), while
spiral galaxies are formed by a relatively continuous star formation.
These models can well reproduce the present gas fractions, colors,
supernova rates, and the color evolution in cluster ellipticals (see
Figures \ref{fig:sgascol} and \ref{fig:ecol}).

Present supernova rates observed in the various type of galaxies
(Cappellaro et al. 1997) put the constraints on the SN Ia progenitor
models.  Using the galaxy model shown in Figures \ref{fig:sgascol} and
\ref{fig:ecol}, we show that our SN Ia model well reproduces the
present supernova rates in both spirals and ellipticals (Kobayashi et
al. 1999).

\subsubsection{Spiral galaxies}

The observed SN II rate ${\cal R}_{\rm II}$ in late-type spirals is about
twice the rate in early-type spirals.  On the other hand, the observed 
SN Ia rate ${\cal R}_{\rm Ia}$ in both types of spirals are nearly the same.
Therefore the present ${\cal R}_{\rm Ia}/{\cal R}_{\rm II}$ ratio in early-type
spirals is about twice that in late-type spirals.  Such a difference in 
the relative frequency is a result of the difference in the SFR, 
because the dependences of ${\cal R}_{\rm II}$ 
and ${\cal R}_{\rm Ia}$ on the SFR are different due to the different 
lifetimes of supernova progenitors. Therefore the observed 
${\cal R}_{\rm Ia}/{\cal R}_{\rm II}$ ratio gives a constraint on the SN Ia 
progenitor model.

In the DD scenario, the lifetime of majority of SNe Ia is $\sim 0.1-0.3$ Gyr, 
so that the evolution of ${\cal R}_{\rm Ia}$ is similar to that of 
${\cal R}_{\rm II}$.  Therefore ${\cal R}_{\rm Ia}/{\cal R}_{\rm II}$ is
insensitive to the SFR.  This results in the small
differences  in ${\cal R}_{\rm Ia}/{\cal R}_{\rm II}$
among the various type of spirals, which is not consistent
with observations.

In our SN Ia model, if the iron abundance of progenitors is
[Fe/H]$\gtsim -1$, the occurrence of SNe Ia is determined from the
lifetime of the companions.  If SNe Ia would occur only in the MS+WD
systems with relatively short lifetimes, ${\cal R}_{\rm Ia}/{\cal R}_{\rm II}$
would have been insensitive to the SFR.  On the contrary, if SNe Ia would occur
only in the RG+WD systems with long lifetimes, the present difference
in ${\cal R}_{\rm Ia}/{\cal R}_{\rm II}$ between early and late type spirals
would have been too large, reflecting the large difference in SFR at an
early epoch.  Owing to the presence of these two types of the
progenitor systems in our SN Ia progenitor model, the observed
difference in ${\cal R}_{\rm Ia}/{\cal R}_{\rm II}$ can be reproduced.

\subsubsection{Elliptical galaxies}

\begin{figure}
\centerline{\psfig{figure=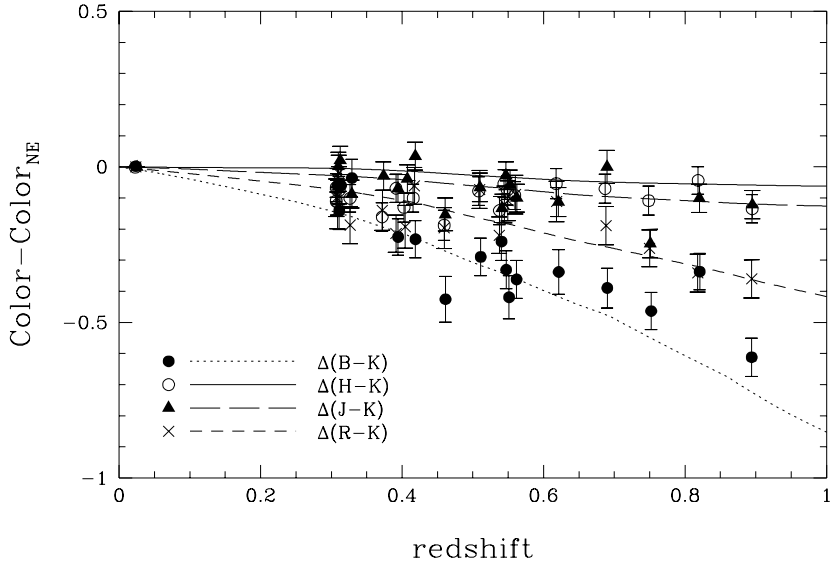,width=12cm}}
\caption[fig4.ps]{\label{fig:ecol}
The passive color evolution of cluster ellipticals up to $z \sim 1$.
The observational data are taken from Stanford et al. (1998).}
\end{figure}

We assume that a bulk of stars in cluster ellipticals are formed at $z
\gsim 3$, and have ages $\gsim 10$ Gyr.  
In the DD scenario, the SN Ia lifetimes are mostly so short
as $t_{\rm Ia}\ltsim 0.3$ Gyr
that too few SNe Ia occur at the present epoch.
Our SN Ia model includes the WD+RG systems with $t_{\rm Ia}\gtsim 10$ Gyr,
thus well reproducing the present SNe Ia rate in ellipticals.

From $z \sim 0.2$ to $z \sim 0$,
the SN Ia rate gradually decreases to the present by $\sim 20\%$.
This is due to the smallest companions with $0.9 M_\odot$ to produce SNe Ia,
which give the longest SN Ia lifetime of
$\sim 11$ Gyr for $Z=0.002$ and $\sim 18$ Gyr for $Z=0.02$.
As the star formation in ellipticals has been stopped
more than 10 Gyr before, majority of SN Ia progenitors with $Z=0.002$
have already explode and only metal-rich SNe Ia occurs at $z \sim 0$.
The decrease in the SN Ia rate from $z \sim 0.2$ to $z \sim 0$ also
depends on the star formation history in ellipticals and the galactic age.
If ellipticals have undergone the relatively continuous star formation,
as suggested by the hierarchical clustering simulations, 
the SN Ia rate might keep constant to the present.  

\subsection{Cosmic supernova rates}

We calculate the cosmic supernova rate by summing up the rates of
spirals and ellipticals with the ratio of the relative mass
contribution among the types.  The relative mass contribution is given
by the observed relative luminosity proportion for ellipticals, 
S0a-Sa, Sab-Sb, Sbc-Sc, and Scd-Sd (\cite{pen76})
and the calculated mass to light ratio in B-band.
We adopt the the initial comoving
density of gas $\Omega_{{\rm g}\infty}=3.5 \times 10^{-3} h^{-1}$.

\subsubsection{In clusters}

\begin{figure}
\centerline{\psfig{figure=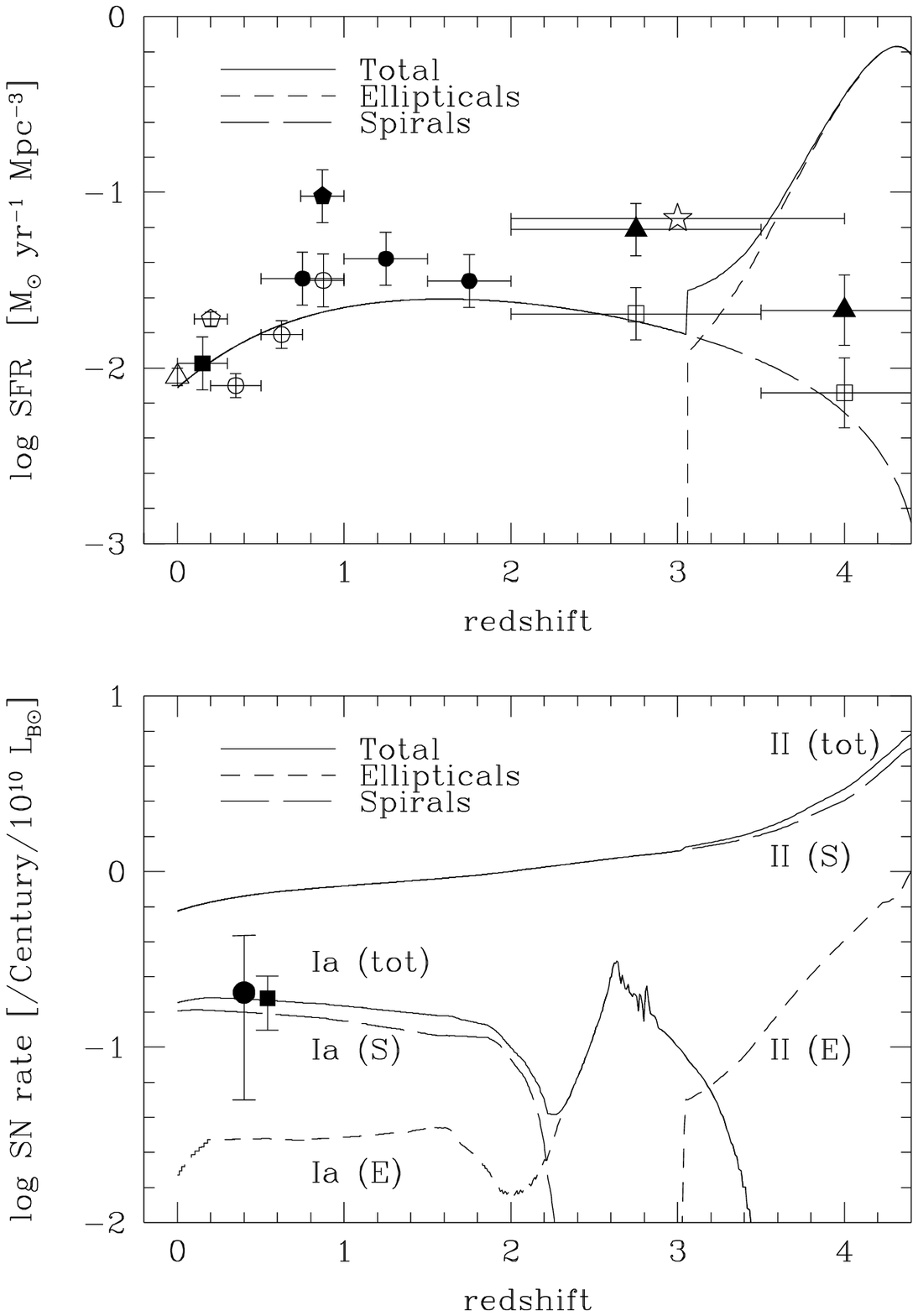,width=15cm}}
\caption[fig6.ps]{\label{fig:esfrsnr}
The upper panel shows the cosmic SFR (solid line) as a composite of 
spirals (long-dashed line) and ellipticals (short-dashed line).
The dots are the observational data (\cite{gal95}, open triangle; 
\cite{lil96}, open circles; \cite{mad96}, open squares; 
\cite{con97}, filled circles; \cite{tres98}, open pentagon; 
\cite{trey98}, filled square; \cite{gla98}, filled pentagon; 
\cite{hug98}, star; \cite{pet98}, filled triangle).
The lower panel shows the cosmic supernova rate (solid line) as a 
composite of spirals (long-dashed line) and ellipticals (short-dashed line).
The observational data are from Pain et al. (1996) and Pain (1999).
}
\end{figure}

First, we make a prediction of the cosmic supernova rates in the
cluster galaxies using the galaxy models which are in good agreements
with the observational constraints (Figs. \ref{fig:sgascol} and
\ref{fig:ecol}).  The upper panel of Figure \ref{fig:esfrsnr} shows 
the cosmic SFR (solid line) as a composite of spirals (long-dashed
line) and ellipticals (short-dashed line).

In our galaxy models, ellipticals undergo a star burst at $z \gtsim 3$
and the duration of the star formation is $\sim 1$ Gyr, while spirals undergo
relatively continuous star formation.  Thus, only the SFR in spirals
is responsible for the cosmic SFR at $z \ltsim 2$.  
For reference, the observed cosmic SFR in field,
so-called Madau's plot,
is also plotted.
Compared with them, the predicted cosmic SFR has a little
shallower slope from the present to the peak at $z \sim 1.4$ (see also
\cite{tot97}).  The high SFR in ellipticals appears at $z \gtsim
3$. Such ellipticals may be hidden by the dust extinction
(\cite{pet98}) or may have formed at $z \sim 5$ (\cite{tot97}).

The chemical enrichment in ellipticals has taken place with so much
shorter timescale that the iron abundance reaches [Fe/H] $\sim -1$ at
$z \gtsim 4$ and the metallicity effect on SNe Ia does not appear.  In
spirals, the iron abundance reaches [Fe/H] $\sim -1$ at $z \sim 2.3$.

The lower panel of Figure \ref{fig:esfrsnr} shows the cosmic supernova
rates (solid line) as a composite of spirals (long-dashed line) and
ellipticals (short-dashed line). The upper and lower three lines show
the SN II and Ia rates, respectively.
Our SN Ia model can successfully reproduce the observed rate at $z \sim 0.5$
(\cite{pai96}, filled circle; \cite{pai99}, filled square).

The SN Ia rate in spirals drops at $z \sim 2$ because of the
low-metallicity inhibition of SNe Ia.  In ellipticals, the chemical
enrichment takes place so early that the metallicity is large enough
to produce SNe Ia at $z \gtsim 3$.  Thus, the SN Ia rate depends almost only
on the lifetime.  A burst of SNe Ia occurs after $\sim 0.5$ Gyr 
from the beginning of the star formation which corresponds to 
the shortest lifetime of the WD+MS binaries; this forms
a peak of the SN Ia rate at $z \sim 2.5$.  The second peak of the SN
Ia rate appears at $z \sim 1.5$ ($\sim$ 2 Gyr) due to the beginning of
the explosions of SNe Ia from WD+RG binaries.  If SNe Ia at $z
\gtsim 2$ are observed with their host galaxies using the Next
Generation Space Telescope, we can precisely test the metallicity
effect by finding the drop of SN Ia rate in spirals.

The cosmic SN Ia rate using the observed cosmic SFR drops at $z \sim
1.6$ ($z \sim 1.2$ in \cite{kob98}).  This is because the chemical
enrichment of the whole universe is much slower than in spirals and
ellipticals.

\subsubsection{In fields}

\begin{figure}
\centerline{\psfig{figure=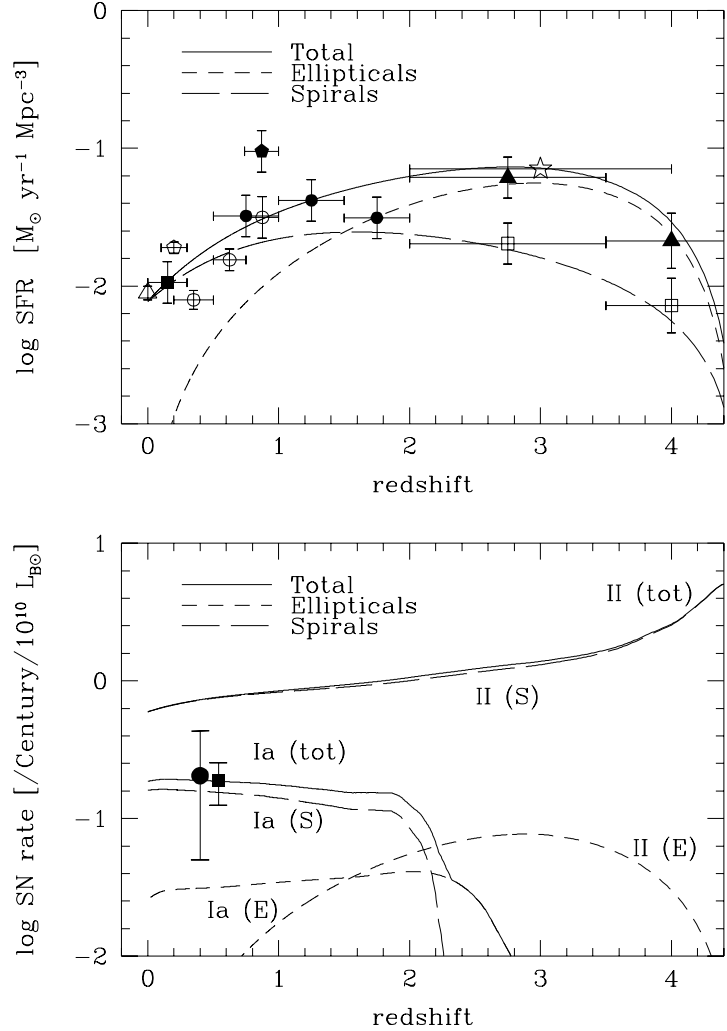,width=15cm}}
\caption[fig9.ps]{\label{fig:esfrsnr2}
The same as Figure \ref{fig:esfrsnr}, but for the formation epochs 
of ellipticals spanning over $1 \lsim z \lsim 4$, 
which might correspond to field ellipticals.}
\end{figure}

The observed spectra of ellipticals in the Hubble Deep Field suggest
that the formation of ellipticals is protracted in fields, that is,
the formation epochs of ellipticals span in the wide range of
redshifts of $1 \lsim z \lsim 4$ (\cite{fra98}).  
We also predict the cosmic supernova rates for this case,
assuming the distribution function of formation epochs
to meet Franceschini's et al. (1998) results.

The upper panel of Figure \ref{fig:esfrsnr2} shows the
cosmic SFR (solid line) as a composite of spirals (long-dashed line)
and field ellipticals (short-dashed line).

The SFR in spirals is the same as in the upper panel of Figure
\ref{fig:esfrsnr}, but the star formation in ellipticals continues to
the present on the average.  The peak of the star formation is at $z
\sim 3$, which is consistent with the recent sub-mm data
(\cite{hug98}).  The synthesized cosmic SFR can successfully
reproduce the observed one, except for the recent H$\alpha$ data
(\cite{gla98}).

The lower panel of Figure \ref{fig:esfrsnr2} shows the cosmic
supernova rates (solid line) as a composite of spirals (long-dashed
line) and ellipticals (short-dashed line).  The upper and lower three
lines show the SN II and Ia rates, respectively.

The SN Ia rate in spirals drops at $z \sim 2$ as in the lower panel of
Figure \ref{fig:esfrsnr}.  Contrary to Figure \ref{fig:esfrsnr}, the
SN Ia rate in field ellipticals gradually decreases from $z \sim 2$ to
$z \sim 3$, which 
is not due to the metallicity effect but due to the lifetime effect.
Although the chemical enrichment timescale in each elliptical
is as fast as in the lower panel of Figure \ref{fig:esfrsnr},
the star formation averaged over whole ellipticals takes place more gradually, 
and the number of stars becomes smaller toward higher redshifts.

In this case, SNe II may be observed in low-redshift ellipticals.
The different SN II and Ia rates between
cluster and field ellipticals reflects the difference in the
star formation histories in different environments.

\subsubsection{Summary}

We make a prediction of the cosmic supernova rate history as a
composite of the different types of galaxies. We adopt the SN Ia
progenitor scenario including the metallicity effect proposed by
Kobayashi et al. (1998), which can successfully reproduce the chemical
evolution of the solar neighborhood.  To calculate the comic SFR, we
construct the galaxy evolution models for spirals and ellipticals to
meet the observational constraints such as the present gas fractions
and colors for spirals, and the mean stellar metallicity and the color
evolution from the present to $z \sim 1$ for ellipticals.

Owing to the {\it two} types of the progenitor system (MS+WD and
RG+WD), i.e., shorter ($0.6-1$ Gyr) and longer lifetimes ($2-15$ Gyr)
of SN Ia progenitors, we can explain the moderate contrast of the
relative ratio of SN Ia to SN II rate ${\cal R}_{\rm Ia}/{\cal R}_{\rm
II}$ between the early and late types of spirals.  Owing to over $10$
Gyr lifetime of the RG+WD systems, SNe Ia can be seen even at the
present in ellipticals where the star formation has already stopped
over 10 Gyr before.

Then we construct the cosmic SFR as the composite of the SFR for
different types of galaxies, and predict the cosmic supernova rates:

\begin{enumerate}
\item 
In the cluster environment, the synthesized cosmic SFR has an excess
at $z \gtsim 3$ corresponding to the SFR in ellipticals and a shallower
slope from the present to the peak at $z \sim 1.4$, compared with
Madau's plot.  The predicted cosmic supernova rate suggests that SNe
Ia can be observed even at high redshifts because the chemical
enrichment takes place so early that the metallicity is large enough
to produce SNe Ia at $z \gtsim 3$ in cluster ellipticals.  In spirals
the SN Ia rate drops at $z \sim 2$ because of the low-metallicity
inhibition of SNe Ia.
\item 
In the field environment, ellipticals are assumed to form at such a
wide range of redshifts as $1 \ltsim z \ltsim 4$.  The synthesized
cosmic SFR has a broad peak around $z \sim 3$, which well reproduces
the observed one.  The SN Ia rate is expected
to be significantly low at $z \gtsim 2$ because the SN Ia rate drops at
$z \sim 2$ in spirals and gradually decreases from $z\sim2$ in ellipticals.
\end{enumerate}

\bigskip

This work has been supported in part by the grant-in-Aid for COE
Scientific Research (07CE2002) of the Ministry of Education, Science,
Culture and Sports in Japan.


\begin{thebibliography}{}

\bibitem[]{}
Arnett, W.D. 1996, Nucleosynthesis and Supernovae
(Princeton: Princeton Univ. Press)

\bibitem[Barbuy \& Erdelyi-Mendes (1989)]{bar89}
Barbuy, B., \& Erdelyi-Mendes, M. 1989, \aap 214 239

\bibitem[]{}
Branch, D. 1998, ARA\&A, 1998, 36, 17

\bibitem[Branch et al. 1983]{bra83}
Branch, D., Lacy, C.H., McCall, M.L., Sutherland, P., Uomoto, A., 
Wheeler, J.C., \& Wills, B.J. 1983, \apj, 270, 123

\bibitem[Branch et al. 1995]{bra95}
Branch, D., Livio, M., Yungelson, L. R., Boffi, F. R., \& Baron, E. 
1995, \pasp, 107, 717

\bibitem[]{}
Branch, D., Romanishin, W., \& Baron, E., 1996, ApJ, 465, 73

\bibitem[Cappellaro et al. 1997]{cap97}
Cappellaro, E., Turatto, M., Tsvetkov, D. Yu., Bartunov, O.S., Pollas,
C., Evans, R., \& Hammuy, M., 1997, \aap, 322, 431

\bibitem[Casoli et al. 1998]{cas98}
Casoli, F., Sauty, S., Gerin, M., Boselli, A., Fouqu\'e, P., Braine,
J., Gavazzi, G., Lequeux, J., \& Dickey, J. 1998, \aap, 331, 451

\bibitem[Connolly et al. 1997]{con97} 
Connolly, A. J., Szalay, A. S., Dickinson, M., SubbaRao, M. U., \&
 Brunner, R. J. 1997, \apj, 486, L11

\bibitem[Cumming et al. 1996]{cum96}
Cumming, R.J., Lundqvist, P., Smith, L.J., Pettini, M., \& King, D.L.
1996, \mnras, 283, 1355

\bibitem[Eck et al. 1995]{eck95}
Eck, C.R., Cowan, J.J., Roberts, D.A., Boffi, F.R., \& Branch, D.
1995, \apj, 451, L53

\bibitem[Edvardsson et al. (1993)]{edv93}
Edvardsson, B., Andersen, J., Gustafsson, B., Lambert, D. L., Nissen, P. E.,
 \& Tomkin, J. 1993, \aap, 275, 101

\bibitem[Franceschini et al. 1998]{fra98}
Franceschini, A., Silva, L., Fasano, G., Granato, G. L., Bressan, A.,
Arnouts, S. \& Danese, L. 1998, \apj, 506, 600

\bibitem[Gallego et al. 1995]{gal95}
Gallego, J., Zamorand, J., Arag\'on-Salamanca, A., \& Rego, M. 1995,
\apj, 455, L1

\bibitem[Garnavich et al. 1998]{gar98}
Garnavich, P. \etal, 1998, ApJ, 493, 53

\bibitem[Gilliland, Nugent \& Phillips 1999]{gil99}
Gilliland, R. L., Nugent, P. E., \& Phillips, M. M. 1999, 
\apj, in press (astro-ph/9903229)

\bibitem[Glazebrook et al. 1999]{gla98}
Glazebrook, K., Blake, C., Economou, F., Lilliy S., \& Colles, M. 1999,
\mnras, 306, 843

\bibitem[Gratton (1991)]{gra91}
Gratton, R. G. 1991, in IAU Symp. 145, 
Evolution of Stars: The Photometric Abundance Connection, 
ed. G. Michaud \& A. V. Tutukov (Montreal: Univ. Montreal), 27

\bibitem[Hachisu \& Kato 1999]{hk99}
Hachisu, I., \& Kato, M., 1999, in preparation

\bibitem[Hachisu, Kato, \& Nomoto 1996]{hac96} 
Hachisu, I., Kato, M., \& Nomoto, K., 1996, ApJ, 470, L97

\bibitem[Hachisu, Kato, \& Nomoto 1999]{hkn99} 
Hachisu, I., Kato, M., \& Nomoto, K. 1999, 
\apj, 522, 487

\bibitem[Hachisu et al. 1999]{hknu99} 
Hachisu, I., Kato, M., Nomoto, K., \& Umeda, H. 1999, 
\apj, 519, 314

\bibitem[]{}
Hamuy, M.,  \etal 1995, AJ, 109, 1

\bibitem[]{}
Hamuy, M.,  Phillips, M. M., Schommer,  R. A., \& Suntzeff, N. B., 
1996, AJ, 112, 2391.

\bibitem[H\"{o}flich \& Khokhlov 1996]{hof96}
H\"oflich, P., \& Khokhlov, A., 1996, ApJ, 457, 500 

\bibitem[]{}
H\"oflich, P., Nomoto, K., Umeda, H., \& Wheeler, J. C., 1999, 
ApJ, in press

\bibitem[]{}
H\"oflich, P., Wheeler, J. C., \& Thielemann, F. -K., 1998, 
ApJ, 495, 617

\bibitem[Hughes et al. 1998]{hug98}
Hughes, D., et al. 1998, Nature, 394, 241

\bibitem[Iben \& Tutukov 1984]{ibe84}
Iben, I. Jr., \& Tutukov, A. V. 1984, \apjs, 54, 335

\bibitem[Iglesias \& Rogers 1993]{igl93}
Iglesias, C. A., \& Rogers, F. 1993, \apj, 412, 752

\bibitem[]{}
Iwamoto, K., Brachwitz, F., Nomoto, K., Kishimoto, N., Umeda, H.,
Hix, W. R., Thielemann, F-K. 1999, ApJS, 125, in press

\bibitem[Kato \& Hachisu 1999]{kat99h}
Kato, M., \& Hachisu, I., 1999, \apj, 513, L41

\bibitem[]{}
Khokhlov, A. 1991, A\&A, 245, 114

\bibitem[Kobayashi \& Arimoto 1999]{kob98b} 
Kobayashi, C., \& Arimoto, N. 1999, \apj, 526, in press

\bibitem[Kobayashi et al. 1999]{kob99} 
Kobayashi, C., Tsujimoto, T., \& Nomoto, K. 1999, \apj, submitted

\bibitem[Kobayashi et al. 1998]{kob98} 
Kobayashi, C., Tsujimoto, T., Nomoto, K., Hachisu, I, \& Kato,
M. 1998, \apj, 503, L155

\bibitem[Kodama \& Arimoto 1997]{kod97} 
Kodama, T., \& Arimoto, N., 1997, \aap, 320, 41

\bibitem[Kodama, Bower \& Bell 1999]{kod98}
Kodama, T., Bower, R. G., \& Bell, E. F. 1999, \mnras, 306, 561

\bibitem[Li \& van den Heuvel 1997]{li97} 
Li, X.-D., \& van den Heuvel, E.P.J., 1997, A\&A, 322, L9

\bibitem[Lilly et al. 1996]{lil96}
Lilly, S. J., Le F\`evre, O., Hammer, F., \& Crampton, D. 1995, \apj,
460, L1

\bibitem[Madau et al. 1996]{mad96} 
Madau, P., Ferguson, H. C., Dickinson, M. E., Giavalisco, M., Steidel,
C. C., \& Fruchter, A. 1996, \mnras, 283, 1388

\bibitem[]{}
Mazzali, P. A., Cappellaro, E., Danziger, I. J., Turatto, M., 
\& Benetti, S., 1998, ApJ, 499, L49

\bibitem[]{}
Niemeyer J.C., \& Hillebrandt W. 1995, ApJ, 452, 769

\bibitem[Nissen et al. (1994)]{nis94}
Nissen, P. E., Gustafsson, B., Edvardsson, B., \& Gilmore, G.
1994, \aap, 285, 440

\bibitem[Nomoto 1982] {nom82}
Nomoto, K., 1982, ApJ, 253, 798

\bibitem[Nomoto, Iwamoto \& Kishimoto 1997a]{nom97}
Nomoto, K., Iwamoto, K., \& Kishimoto, N. 1997a, Science, 276, 1378

\bibitem[Nomoto et al. 1997b]{nom97b}
Nomoto, K., Hashimoto, M, Tsujimoto, T, Thielemann, F.-K, Kishimoto,
N., Kubo, Y., \& Nakasato, N. 1997b, Nuclear Physics, A616, 79c

\bibitem[Nomoto et al. 1997c]{nom97c}
Nomoto, K., Iwamoto, K., Nakasato, N., Thielemann, F.-K, Brachwitz,
F., Tsujimoto, T., Kubo, Y., \& Kishimoto, N. 1997c, Nuclear Physics,
A621, 467c

\bibitem[]{}
Nomoto, K., et al. 1997d, in Thermonuclear Supernovae, Eds.
 P.Ruiz-Lapuente et al. (Dordrecht: Kluwer), 349

\bibitem[Nomoto \& Kondo 1991]{nom91}
Nomoto, K., \& Kondo, Y., 1991, ApJ, 367, l19

\bibitem[]{}
Nomoto, K., Nariai, K., \& Sugimoto, D., 1979, PASJ, 31, 287

\bibitem[]{}
Nomoto, K., Thielemann, F. -K., \& Yokoi, K., 1984, ApJ, 286, 644 

\bibitem[Nomoto et al. 1994]{nom94}
Nomoto, K., Yamaoka, H., Shigeyama, T., Kumagai, S., \& Tsujimoto, T.
1994, in Supernovae, Les Houches Session LIV, ed. S. A. Bludman et al.
(Amsterdam: North-Holland), 199

\bibitem[Nugent et al. 1997]{nug97}
Nugent, P., Baron, E., Branch, D., Fisher, A., \& Hauschildt,
P. H. 1997, \apj, 485, 812

\bibitem[]{}
Paczy\'nski, B. 1973, Acta Astr. 23, 1

\bibitem[Pain 1999]{pai99} 
Pain, R. 1999, Talk at the Type Ia supernova workshop in Aspen

\bibitem[Pain et al. 1996]{pai96} 
Pain, R., et al. 1996, \apj, 473, 356

\bibitem[Pei \& Fall 1995]{pei95} 
Pei, Y. C., \& Fall, S. M., 1995, \apj, 454, 69

\bibitem[Pei, Fall \& Hauser 1999]{pei98}
Pei, Y. C., Fall, S. M., \& Hauser, M. G. 1999, \apj, in press

\bibitem[Pence 1976]{pen76}
Pence, W. 1976, \apj, 420, L1

\bibitem[Perlmutter et al. 1997]{per97}
Perlmutter, S. et al. 1997, \apj, 483, 565

\bibitem[]{}
Perlmutter, S. \etal 1999, ApJ, 517, 565

\bibitem[Pettini et al. 1998]{pet98}
Pettini, M., Kellogg, M.,Steidel, C. C., Dickinson, M., Adelberger, K. 
L, \& Giavalisco, M. 1998, \apj, 508, 539

\bibitem[]{}
Phillips,  M. M., 1993, ApJ, 413, L75

\bibitem[]{}
Riess,  A.G., Press, W.H., \& Kirshner, R.P. 1995, ApJ, 438, L17

\bibitem[]{}
Riess,  A.G. \etal 1998, AJ, 116, 1009

\bibitem[]{}
Riess,  A. G. \etal 1999, AJ, 117, 707

\bibitem[Roberts \& Haynes 1994]{rob94}
Roberts M. S. \& Haynes M. P. 1994, \araa, 32, 115

\bibitem[Ruiz-Lapuente \& Canal 1998]{rui98}
Ruiz-Lapuente, P., \& Canal, R. 1998, \apj, 497, L57

\bibitem[Sadat et al. 1998]{sad98}
Sadat, R., Blanchard, A., Guiderdoni, B., \& Silk, J. 1998, \aap, 331,
L69

\bibitem[Saio \& Nomoto 1985]{sai85}
Saio, H., \& Nomoto, K. 1985, \aap, 150, L21

\bibitem[Saio \& Nomoto 1998]{sai98}
Saio, H., \& Nomoto, K. 1998, \apj, 500, 388

\bibitem[]{}
Sandage, A., \& Tammann, G.A., 1993, ApJ, 415, 1

\bibitem[Schlegel \& Petre 1993]{sch93}
Schlegel, E.M., \& Petre, R. 1993, \apj, 418, L53

\bibitem[Segretain et al. 1997]{seg97}
Segretain, L., Chabrier, G., \& Mochkovitch, R. 1997, \apj, 481, 355

\bibitem[Stanford et al. 1998]{sta98} 
Stanford, S.A., Eisenhardt, P.R.M., \& Dickinson, M., 1998, \apj, 492,
461

\bibitem[Suzuki \& Nomoto 1995]{suz95}
Suzuki, T., \& Nomoto, K. 1995, \apj, 455, 658

\bibitem[Totani et al. 1997]{tot97}
Totani, T., Yoshii, Y., \& Sato, K. 1997, \apj, 483, L75

\bibitem[Tresse \& Maddox 1998]{tres98}
Tresse, L., \& Maddox, S. J. 1998, \apj, 495, 691

\bibitem[Treyer et al. 1998]{trey98}
Treyer, M. A., Ellis, R. S., Milliard, B., Donas, J., \& Bridges,
T. J. 1998, \mnras, 300, 303

\bibitem[Tsujimoto et al. 1995]{tsu95}
Tsujimoto, T., Nomoto, K., Yoshii, Y., Hashimoto, M., Yanagida, S., \&
Thielemann, F.-K. 1995, \mnras, 277, 945

\bibitem[Tutukov \& Yungelson 1994]{tut94} 
Tutukov, A. V., \& Yungelson, L. R. 1994, \mnras, 268, 871

\bibitem[Umeda et al. 1999a]{ume99a}
Umeda, H., Nomoto, K., Ymamaoka, H., \& Wanajo, S. 1999a, \apj, 513,
861

\bibitem[Umeda et al. 1999b]{ume99b}
Umeda, H., Nomoto, K., Kobayashi, C., Hachisu, I, \& Kato, M. 1999b,
ApJL, 522, in press (astro-ph/9906192)

\bibitem[]{}
van den Heuvel, E.P.J., 1994, in Interacting Binaries,
eds. S.N. Shore \etal (Berlin: Springer-Verlag), 263

\bibitem[van den Heuvel et al. 1992]{heu92}
van den Heuvel, E. P. J., Bhattacharya, D., Nomoto, K., \& Rappaport, 
S. 1992, \aap, 262, 97  

\bibitem[]{}
Wang, L., H\"oflich, P., \& Wheeler, J. C. 1997, ApJ, 483, L29

\bibitem[Webbink 1984]{web84}
Webbink, R. F. 1984, \apj, 277, 355

\bibitem[Yungelson \& Livio 1998]{yun98}
Yungelson, L., \& Livio, M. 1998, \apj, 497, 168

\bibitem[]{}
Zaritsky, D., Kennicutt, R.C., \& Huchra, J.P. 1994, ApJ, 420, 87

\end{thebibliography}
\end{document}